 \documentclass[twocol]{ametsoc}

 \journal{jpo}

\usepackage{verbatim}

\title{Surface expression of a wall fountain: application to subglacial discharge plumes}

\authors{Craig D. McConnochie\correspondingauthor{Craig Mcconnochie, Department of Civil and Natural Resources Engineering, The University of Canterbury, Christchurch, New Zealand.}}

\affiliation{Department of Civil and Natural Resources Engineering, The University of Canterbury, Christchurch, New Zealand}

\email{craig.mcconnochie@canterbury.ac.nz}

\extraauthor{Claudia Cenedese}
\extraaffil{Physical Oceanography Department, Woods Hole Oceanographic Institution, Woods Hole, MA, USA}

\extraauthor{Jim N. McElwaine}
\extraaffil{Department of Earth Sciences, University of Durham, Durham, UK\\
Planetary Science Institute, Tucson, AZ, USA}

\abstract{We use laboratory experiments and theoretical modeling to investigate the surface expression of a subglacial discharge plume, as occurs at many fjords around Greenland.
The experiments consider a fountain that is released vertically into a homogeneous fluid, adjacent either to a vertical or a sloping wall, that then spreads horizontally at the free surface before sinking back to the bottom.
We present a model that separates the fountain into two separate regions: a vertical fountain and a horizontal, negatively buoyant jet. 
The model is compared to laboratory experiments that are conducted over a range of volume fluxes, density differences, and ambient fluid depths. 
It is shown that the non-dimensionalised length, width, and aspect ratio of the surface expression are dependent on the Froude number, calculated at the start of the negatively buoyant jet. 
The model is applied to observations of the surface expression from a Greenland subglacial discharge plume.
In the case where the discharge plume reaches the surface with negative buoyancy the model can be used to estimate the discharge properties at the base of the glacier.}

\begin{document}

\maketitle

%








\section{Introduction}

The mass loss from the Greenland and Antarctic ice sheets is an increasingly significant component of global sea level rise \citep[e.g.][]{Chambers17}. 
Interactions between the polar oceans and the ice sheets are an important control on the rate of glacier melting, but are not fully understood \citep{Straneo15}. 
An area of recent focus has been the subglacial discharge plumes that are released at the base of Greenland glaciers and that have been linked to elevated melt-rates \citep[e.g.][]{Slater16,Carroll15,Straneo15}.

These subglacial discharge plumes originate from surface melting of the ice sheet.
The meltwater then flows to the base of the ice sheet where it travels underneath the ice through a complex network of channels. 
Eventually, the meltwater is released into the fjord at the grounding line --- the location where the ice sheet becomes afloat.
There is currently a high degree of uncertainty regarding the geometry of the subglacial discharge sources with possibilities ranging from relatively confined point sources to extended line sources \citep{Jackson17}.
The fresh ($\approx 0\,{\rm g/kg}$) and cold ($\approx 0 {\rm {}^oC}$) meltwater is lighter than the surrounding fluid and forms a turbulent plume that rises up the ice face entraining relatively warm and salty water from the fjord. 
The entrained fjord water rapidly mixes throughout the plume, enhances the transport of heat and salt to the ice boundary and drives rapid melting, as demonstrated in laboratory experiments by \cite{McConnochie17} and \cite{Cenedese16b}.
Here we focus on glaciers that have vertical or near-vertical termini as opposed to glaciers with a near-horizontal floating ice shelf or ice tongue.
With this focus, it is typically assumed that the ice face near subglacial plumes is vertical although recent observations have suggested that the ice face can be undercut \citep{Fried15}.

Greenland fjords typically have an approximately two-layer stratification \citep{Straneo11,Straneo12}. 
Due to this density stratification, subglacial plumes often rise to a level where they are denser than the ambient fluid.
At this point, the flow is best described as a fountain that rises due to its momentum but decelerates due to its buoyancy.
The flow continues to entrain relatively warm and salty water from the fjord but instead of accelerating it begins to decelerate.
Throughout this paper we typically use the term plume to describe the general flow resulting from the subglacial discharge of meltwater and the term fountain to describe the specific part of the flow containing upward vertical momentum and downward buoyancy.
After rising vertically, the meltwater plume can either intrude horizontally into the ambient stratification at mid-depth if its density is higher than the upper layer, propagate away from the ice at the free surface if its density is lower than the upper layer, or reach the free surface due to excessive vertical momentum before sinking back down to a level of neutral buoyancy \citep{Beaird15,Cenedese16b,Mankoff16,Sciascia13}.

In this study we focus on the third scenario where the meltwater rises through the upper layer as a fountain and then sinks. 
In contrast to canonical fountains in semi-infinite environments where the flow rises until it has zero vertical momentum and then falls back around the rising fluid to a level of neutral buoyancy \citep{Hunt15}, we focus here on flows that reach the free surface with significant vertical momentum. 
This causes the fluid to spread horizontally at the surface for some distance before sinking to a level of neutral buoyancy. 
Since the subglacial discharge is typically highly turbid, it is often visible as a pool of sediment-laden fluid at the free surface \citep[e.g.][]{How17,Mankoff16}.
Although the suspended sediment could have important dynamical effects on the plume and the surface pool, throughout this study we will assume that it is carried as a passive tracer.

\cite{Mankoff16} photographed a well defined region of turbid fluid in front of Saqqarliup Sermia, Greenland, that was interpreted as a subglacial discharge \emph{pool}. 
The pool was triangular in shape, approximately 300\,m wide at the glacier face and stretched 300\,m away from the glacier. 
Considering that subglacial discharge plumes are typically assumed to be semi-circular \citep{Slater16,Mankoff16}, the triangular shape is somewhat surprising and as yet, has not been explained.

There are several possible explanations for the triangular shaped surface expression such as a sloping glacier face, a secondary circulation induced by the narrow fjord, and increased melting induced by the plume itself leading to an incised glacier face that redirects the surface outflow. 
In this paper we use a set of laboratory experiments to investigate several controlling mechanisms on the surface expression of subglacial discharge plumes which we model as a fountain, next to a wall, that reaches the free surface. 
We consider a fountain rather than a plume as we are interested in a flow that will reach the free surface, spread for some distance and then sink. 
By considering a fountain we effectively limit the investigation to the region where the flow is above its level of neutral buoyancy.

Although both the vertical plume \citep[e.g.][]{Slater16} and horizontal intrusion \citep[e.g.][]{Jackson17} have been extensively studied, much less attention has been given to how the flow transitions from vertical to horizontal.
This transition is important not just for determining the size of the surface expression, but also for understanding how to force the boundary of fjord-scale models \citep[e.g.][]{Cowton15,Carroll15}.
As such, although this study is primarily focused on the size of the surface expression, it will also offer insight into how the subglacial discharge flow transitions from vertical to horizontal which has broader implications for Greenland fjords.

Turbulent fountains have been extensively studied in the past \citep[see][]{Hunt15}. 
Many of the previous studies have examined the entrainment of ambient fluid into an axisymmetric fountain \citep[e.g.][]{Burridge16,Bloomfield98}. 
Although turbulent fountains and turbulent plumes are governed by approximately the same force balance, entrainment into fountains is significantly more complex due to the
potential re-entrainment of sinking fluid into the rising fountain. 
The problem of a fountain adjacent to a wall has also been of interest to many authors given its applicability to building fires and enclosed convection \citep[e.g.][]{Goldman86,Kapoor89,Kaye07}.

Despite the previous work on turbulent fountains, there are features of subglacial discharge fountains that have not been fully studied. 
First, much of the previous work on wall fountains has focused on two-dimensional flows whereas we are interested in wall fountains generated from a point source. 
In addition, subglacial discharge fountains reach the free surface with a significant vertical momentum that causes them to spread horizontally before sinking. 
As such the upward and downward flows can be spatially separated, causing the horizontal flow field at the free surface to be important to the overall dynamics. 
As well as the applicability to subglacial discharge surface expressions, understanding this surface flow could be important in a variety of similar problems as it will control where the source fluid will come to rest and how much entrainment of ambient fluid will occur.

From the definition of a fountain given in \cite{Hunt15}, the fact that the sinking fluid is not re-entrained into the rising flow would suggest that the considered situation is not in fact a fountain but is a vertical, negatively buoyant jet.
However, we retain the term fountain for simplicity and to clarify that the rising flow is denser than the surrounding fluid, unlike what would be expected for a plume.

The purpose of this paper is to investigate some of the physical processes that could control the dimensions and shape of the surface expression of a subglacial discharge plume. 
Our hypothesis is that if the processes controlling the surface expression are well understood, then the subglacial discharge properties could be inferred from visual observations of the fjord surface.

In \S\ref{sec:Theory} we present a theoretical model of the fountain that results from a subglacial discharge plume and of its surface expression. 
In \S\ref{sec:Experiments} we describe the experimental apparatus. 
\S\ref{sec:Spreading} and \S\ref{sec:Surface} give the experimental results and comparisons with predictions from the theoretical model. 
Finally in \S\ref{sec:Application} we apply the model to the observations of a surface pool described in \cite{Mankoff16} and attempt to infer the sub-surface properties of the plume.

\section{Theory}
\label{sec:Theory}
We consider a scenario similar to that occurring at the Greenland glacier fronts where floating ice shelves are not present: the steady and vertical release of freshwater, from a single point source located next to a wall, into a relatively deep (many source radii) two-layer stratification. 
The release of freshwater will produce buoyant fluid that is often modelled as a semi-circular plume \citep{Slater16,Mankoff16}. 
Although the source of the freshwater discharge is likely to have some horizontal momentum in the geophysical case, it is common to model the discharge as a purely vertical flow as the length scale whereby the discharge attaches to the wall is typically much smaller than the total water depth \citep[e.g.][]{Cowton15,Xu13}.

\begin{figure}
\centering
  \noindent\includegraphics[width=8cm]{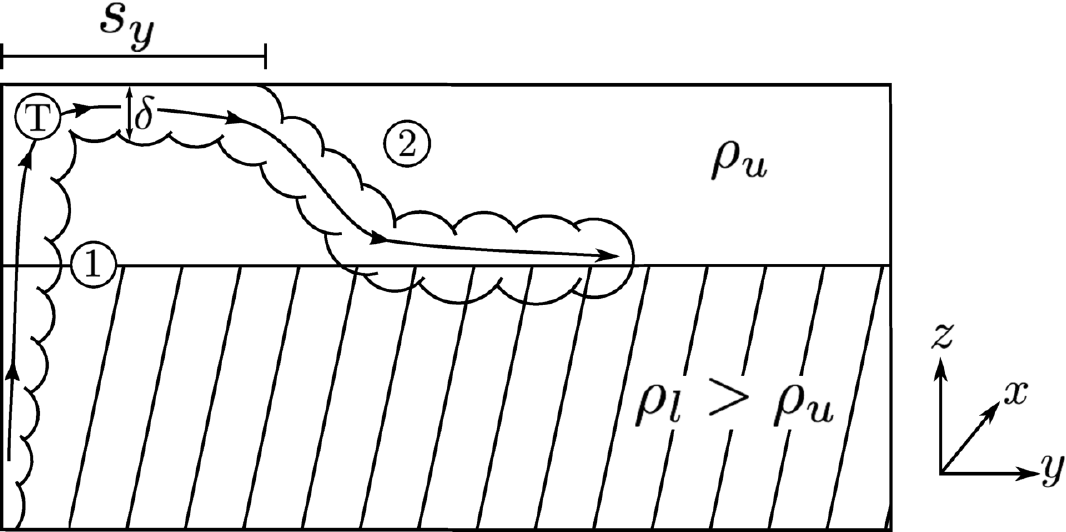}
  \caption{Schematic showing the flow being considered. 
  Relatively fresh fluid is released into a two-layer stratification with upper layer density $\rho_u$ and lower layer density $\rho_l$. 
  Labels 1, 2, and T show the two regions of the flow and the transition from vertical to horizontal flow. 
  The experiments only consider the upper (unhatched) layer. 
  Therefore the interface between the two density layers on this figure is the base of the experimental tank. 
  $s_y$ shows the length of the surface expression in the $y$ (wall-perpendicular) direction as measured in our experiments and $\delta$ is the thickness of the horizontal negatively buoyant jet.}
  \label{fig:schematic}
\end{figure}

To produce a surface expression of the plume that is denser than the upper layer, we set the ambient density profile and subglacial discharge characteristics such that the plume is initially positively buoyant but, due to entrainment of the lower layer fluid, it becomes more dense than the upper layer. 
However, the vertical momentum at the interface between the lower and upper layers is assumed to be sufficient that the plume will rise to the free surface and spread horizontally for some distance before sinking back to the interface depth. 
This is shown schematically in figure~\ref{fig:schematic}.

The flow can be considered in two separate regions: 
first, a vertical flow next to the wall and 
second, a horizontal negatively buoyant jet that is generated at the free surface. 
These regions are shown on figure~\ref{fig:schematic} and shall be referred to as ``region 1" and ``region 2".
As the flow transitions from the first to the second region we assume that all vertical momentum is converted to horizontal momentum. 
In the appendix we show how the vertical momentum is converted into horizontal momentum in the transition region. 
Details of the two separate regions and the transition are described below.

To simplify the experiments (\S\ref{sec:Experiments}), we will ignore the positively buoyant plume in the lower layer and consider a dynamically equivalent system only comprising the lighter upper layer. 
The initial discharge in the simplified system is equivalent to the plume at the density interface in the full geophysical system.
The experimental system is shown as the unhatched region in figure~\ref{fig:schematic}.
In the following section we consider only the simplified system but note that the same equations that are used for the fountain in the upper layer can be applied to the buoyant plume that forms in the lower layer.

\subsection{Wall fountain}
\label{sec:phase1}
The canonical equations for a buoyant plume from \cite{Morton56} are used for both the positively buoyant flow (the plume) in the lower layer and the negatively buoyant flow (the fountain) in the upper layer of region 1. 
Following \citet{Slater16} and \citet{Ezhova18} we have assumed that the wall causes the plume to be semi-circular and adjusted the plume model accordingly.
The fountain volume flux $Q$, momentum flux $M$, and buoyancy flux $B$ are calculated from
\begin{eqnarray}
  \frac{{\rm{d}}Q}{{\rm{d}}z} &=& \pi \alpha b w,       \label{MTT_Q}\\
  \frac{{\rm{d}}M}{{\rm{d}}z} &=& \frac{\pi b^2 g'}{2} - 2C_d b w^2,    \label{MTT_M}
\end{eqnarray}
and
\begin{eqnarray}
  \frac{{\rm{d}}B}{{\rm{d}}z} &=& \frac{{\rm{d}}}{{\rm{d}}z}\left(\frac{\pi b^2 w g'}{2}\right) = 0     \label{MTT_B}
\end{eqnarray}
where 
$z$ is the height above the source, 
$b$ is the top-hat fountain radius, 
$w$ is the top-hat fountain velocity, 
$g'$ is the top-hat reduced gravity between the fountain and the surrounding ambient fluid, 
$C_d$ is the drag coefficient assumed to be 0.0025 \citep{Cowton15} 
and $\alpha=w_e/w$ is the entrainment coefficient with $w_e$ being the velocity with which ambient fluid is entrained into the fountain. 
We note that $ M $ is actually the specific momentum flux (i.e. the momentum flux divided by the density) but will be referred to as the momentum flux throughout the paper for simplicity.
The value for $\alpha$ is determined to be 0.10 by the experiments described in \S\ref{sec:Spreading}. 
It is assumed that the drag against the wall is negligible compared to the buoyancy forces. 
Using a drag coefficient of 0.0025, the drag force is estimated to be approximately 5\% of the buoyancy force in the laboratory experiments and typically $<3\%$ of the buoyancy force for a geophysical scenario. 

Equations (\ref{MTT_Q})--(\ref{MTT_B}) are initialised at $z=0$ (the base of the unhatched layer on figure~\ref{fig:schematic}) using the values of the volume, momentum and buoyancy fluxes and the discharge area at the source. 
As demonstrated by \cite{Kaye06}, for (\ref{MTT_Q})--(\ref{MTT_B}) to be valid the fountain must have a source Froude number ($ {\rm Fr_0} = w_0(g_0'b_0)^{-1/2} $) greater than approximately 3, where subscript $0$ refers to source properties. 
The width of the plume must also be much less than the thickness of the top layer.
In the context of a subglacial discharge, these conditions will generally be met in cases where the flow reaches the surface.
Equations (\ref{MTT_Q})--(\ref{MTT_B}) are used to determine the vertical fluxes at the free surface.

Following \cite{Ezhova18}, we expect the time-averaged density and velocity profiles in the wall-parallel direction to be roughly Gaussian.
Perpendicular to the wall, the velocity and density profiles outside of the viscous sublayer can also be adequately modelled by a Gaussian curve, although the rate of spread is significantly lower in the wall-perpendicular than the wall-parallel direction.
The more rapid spreading in the wall-parallel direction is explained by \cite{Launder83} in the context of wall jets by the interaction with the wall leading to non-isotropic turbulent fluctuations --- eddies normal to the wall cannot be as large as eddies parallel to the wall.

At the level of the free surface, we expect the fountain to have a density maximum  located at the wall but a velocity maximum that is offset some distance due to the no-slip boundary condition imposed by the wall. 
The distance of this offset, $y_{\rm o}$, is taken from direct numerical simulations of a wall plume \citep{Ezhova18} and we approximate the velocity profile between the wall and the maximum velocity location as linear in the $y$ direction and Gaussian in the $x$ direction. 
The velocity and density profiles at the height of the undisturbed free surface can then be described as
\begin{equation}
  w_s(x,y) = \left\{
    \begin{array} {ll}
      \overline{w} \exp\left[-\frac{1}{2}\left( \frac{x^2}{m^2}  + \frac{(y-y_{\rm o})^2}{n^2} \right)\right], & y>y_{\rm o} \\[14pt]
      \left(\frac{y\,\overline{w}}{y_{\rm o}}\right)\exp\left[-\frac{x^2}{2m^2}\right],         & y<y_{\rm o}
   \end{array} 
\right.
  \label{eq:wxy}
\end{equation}
and
\begin{equation}
  g'_s = \overline{g'} \exp\left[-\frac{1}{2}\left(\frac{x^2}{m^2}+\frac{y^2}{n^2}\right)\right],
  \label{eq:gxy}
\end{equation}
where $\overline{w}$ and $\overline{g'}$ are the maximum values of the fountain velocity and reduced gravity, $x$ and $y$ are the distances in the wall-parallel and wall-perpendicular direction, 
and $m$ and $n$ define the size of the fountain in the $x$ and $y$ directions, respectively. 
The values of $\overline{w}$ and $ \overline{g'} $ are taken from the free surface values of equations~(\ref{MTT_Q})--(\ref{MTT_B}) and depend on the source conditions and ambient fluid properties (i.e. depth and density structure). 
The values of $m$ and $n$ are taken from experiments that measured the width and the length of the fountain before it reached the surface (\S\ref{sec:Spreading}). 
Finally, $x$ and $y$ are defined such that $(x,y)=(0,0)$ gives the centre of the fountain in the $x$ direction and the position of the wall in the $y$ direction. 
The exact form of the velocity profile for the region $y<y_{\rm o}$ is relatively unimportant to the spreading of the surface expression. 
The fluid between $y_{\rm o}$ and the wall spreads parallel to the wall rather than away from the wall, so it has almost no effect on the length of the surface expression (the size in the wall-perpendicular direction). 
The velocity profile has only a small effect on the width of the surface expression (the size in the wall-parallel direction) due to the small fluxes in this region. 
As the Reynolds number increases the region between $y=0$ and $y = y_{\rm o}$ will contain a smaller proportion of the total fluxes so in a geophysical setting, with a very large Reynolds number, this region is most likely completely insignificant.

\subsection{Transition from vertical to horizontal fluxes}
\label{sec:transition}
We assume that the fountain's momentum causes the free surface to rise a small amount so that the flow can be treated as equivalent to the solution for the flow around a $90^\circ$ corner.
This assumption is justified in the appendix. 
The pressure at the free surface $p_s$ leads to the free surface rising according to
\begin{equation}
  {{Z}}(x,y) = \frac{p_s(x)}{g} = \frac{w_s^2-u_s(x)^2/2}{g},
  \label{eq:Surface_deviation}
\end{equation}
where $g$ corresponds to the acceleration due to gravity and not the reduced gravity.
$u_s(x)$ is the horizontal velocity at the surface and the second part of this equation follows from Bernoulli's principle.

\begin{figure*}
\centering
\noindent\includegraphics[width=39pc]{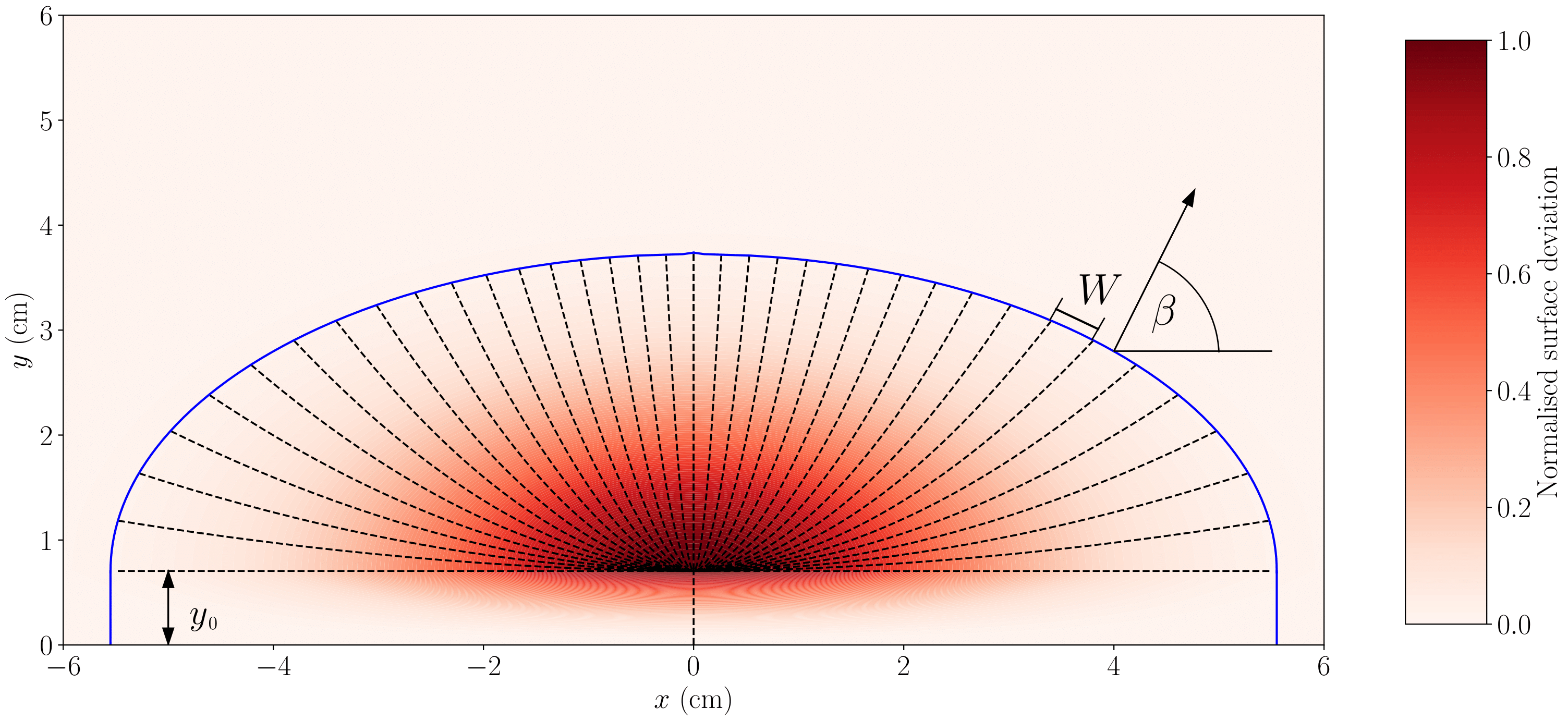}
\caption{The modelled free surface deviation caused by the fountain impacting the free surface. 
The blue line shows the defined edge of the fountain and the black dashed lines show the sector boundaries. 
$\beta$ is the angle normal to the fountain edge that the jet will propagate in. 
$W$ is the width of the sector. 
In this case, the value of $y_{\rm o}$ is 0.71\,cm.}
\label{fig:Surface_deviation}
\end{figure*}

Figure~\ref{fig:Surface_deviation} shows the modelled free surface deviation {$Z$}, normalised by the maximum value. 
A threshold free surface deviation of ${{Z}} = 0.01{{Z}}_{\rm{max}}$ is used to define the outside edge of the fountain (blue line on figure~\ref{fig:Surface_deviation}).
Following \citet{Zgheib15}, we separate the fountain into independent sectors. 
The sectors are defined such that the arc angle that the sector boundaries make with the centre of the fountain is constant and that the sector edges follow the maximum gradient in the free surface (black dashed lines on figure~\ref{fig:Surface_deviation}). 
As such, all sectors start from the centre of the fountain, follow the steepest gradient of the free surface and finish at the fountain boundary at uniformly spaced angles. 
Due to the asymmetric Gaussian velocity profiles, this results in sector boundaries that are slightly curved rather than being straight lines as in \citet{Zgheib15}. 
All of the fluid between $y_{\rm o}$ and the wall will travel in the wall-parallel direction so this region is treated as two sectors: one in the positive $x$ direction and the other in the negative $x$ direction.

The vertical volume, momentum and buoyancy fluxes entering each sector are calculated from the velocity and reduced gravity profiles given in~(\ref{eq:wxy}) and~(\ref{eq:gxy}) as
\begin{eqnarray}
  \hat{Q} &=& \oint\limits_S w \,{\rm d} A,\\
  \hat{M} &=& \oint\limits_S w^2 \,{\rm d} A,\\
  \hat{B} &=& \oint\limits_S wg' \,{\rm d} A,
\end{eqnarray}
where $\hat{\cdot}$ refers to the vertical flux in a sector and
$\oint\limits_S {\rm d} A$ is the surface integral over the area of a sector.
The fluxes are combined with the sector width $W$ (see figure~\ref{fig:Surface_deviation}) to calculate the top-hat velocity $u$, reduced gravity $g'$, and thickness $\delta$
of the negatively buoyant jets that leave the sector horizontally:
\begin{eqnarray}
  u &=& \frac{\hat{M}}{\hat{Q}},\\
  g' &=& \frac{\hat{B}}{\hat{Q}},\\
  \delta &=& \frac{\hat{Q}^2}{\hat{M}W}.
  \label{jet_h}
\end{eqnarray}
The horizontal velocity $u$ leaving a given sector is assumed to be in the direction of maximum free-surface gradient at the centre of the sector as shown by the arrow on figure~\ref{fig:Surface_deviation} and consistent with the definition of the sector boundaries.

\subsection{Horizontal negatively buoyant jet}
\label{sec:phase3}

The second region is composed of a series of negatively buoyant jets, directed horizontally, and emanating from the outside boundary of the sectors shown in figure~\ref{fig:Surface_deviation}. 
The velocity $u$, reduced gravity $g'$, thickness $\delta$, width $W$, and direction $\beta$ of the jet are all obtained from \S\ref{sec:Theory}\ref{sec:transition}. 
It is envisaged that a separate jet is leaving from each sector. 
The jets have a cross-sectional area given by the thickness $\delta$ and the sector width $W$ and are bounded on the top by the free surface, on either side by the neighbouring jets (or the wall), and ambient fluid on the base. 
As such, once the surface expression has reached a quasi-steady size, the jets only entrain ambient fluid through the base. 
Once the surface expression begins to sink, the jets will be able to entrain ambient fluid from above as well but this is outside the focus of our model.
The equations that govern the propagation of each jet are similar to those given in (\ref{MTT_Q})--(\ref{MTT_B}) but are adapted for the different geometry and orientation:
\begin{eqnarray}
  \frac{{\rm{d}}Q}{{\rm{d}}\bf{s}} &=& \alpha uW,    \label{jet_Q}\\
  \frac{{\rm{d}}M_{x,y}}{{\rm{d}}\bf{s}} &=& 0, \\
  \frac{{\rm{d}}M_z}{{\rm{d}}\bf{s}} &=& W \delta g',
\end{eqnarray}
and
\begin{eqnarray}
  \frac{{\rm{d}}B}{{\rm{d}}\bf{s}} &=& 0,    \label{jet_B}
\end{eqnarray}
where $\bf{s}$ is the distance along the path length of each jet, 
and $M_{x,y}$ and $M_z$ are the momentum fluxes in the horizontal and vertical directions, respectively. 
An entrainment coefficient of $\alpha=0.1$ is used in this region, which is a value between that of a pure jet and a pure plume \citep{Carazzo16}.

The jet centreline position, ${\bf{s}} = (s_x,s_y,s_z)$, is also tracked for each sector over time as:
\begin{eqnarray}
  \frac{{\rm{d}}s_x}{{\rm{d}}{\bf{s}}} &=& \cos\beta  \, \cos \left[\tan^{-1}\left(\frac{M_z}{M_{x,y}}\right)\right],
  \label{sx}\\
  \frac{{\rm{d}}s_y}{{\rm{d}}{\bf{s}}} &=& \sin\beta  \, \cos \left[\tan^{-1}\left(\frac{M_z}{M_{x,y}}\right)\right],
  \label{sy}\\
  \frac{{\rm{d}}s_z}{{\rm{d}}{\bf{s}}} &=& -\sin \left[\tan^{-1}\left(\frac{M_z}{M_{x,y}}\right)\right],
  \label{sz}
\end{eqnarray}
where $\beta$ is the horizontal angle of jet propagation taken from \S\ref{sec:Theory}\ref{sec:transition} and measured from a plane that is parallel to the wall, as shown on figure~\ref{fig:Surface_deviation}.
The value $\gamma = \left(\tan^{-1}\left({M_z}/{M_{x,y}}\right)\right)$ gives the angle of jet propagation in the vertical plane, measured from the horizontal and increasing downwards 
(i.e. $\gamma$ is initially zero and increases as the horizontal jet becomes a vertical plume).
Finally, $s_z$ is constrained such that the distance between the jet centreline and the free surface is at least half the jet thickness.
\mbox{Equations~(\ref{jet_Q})--(\ref{sz})} are evolved in $\bf{s}$ from the edge of the fountain (blue line on figure~\ref{fig:Surface_deviation}) until the upper surface of the jet is deeper than a predetermined threshold based on the experimental setup, at which point $s_x$ and $s_y$ define the outside edge of the fountain surface expression for each sector.

Different sectors have different momentum and buoyancy fluxes due to both the fountain asymmetry and the offset between the Gaussian velocity and density profiles. 
The fountain asymmetry results in the edge of the fountain (blue line on figure~\ref{fig:Surface_deviation}) having a larger radius of curvature in the centre \mbox{($x\approx 0$)} than near the wall \mbox{($y\rightarrow y_{\rm o}$)}. 
This causes the jet to decelerate more rapidly in the wall-parallel direction than in the wall-perpendicular direction. 
The offset in the velocity and density profiles results in the sectors in the centre of the fountain having a lower reduced gravity and higher velocity than the sectors near the wall. 
Both of these factors will cause the surface expression to spread further in the wall-perpendicular direction than in the wall-parallel direction, as will be shown when we compare the model predictions with the experimental observations (figure~\ref{fig:plume_surface}).

\section{Experiments}
\label{sec:Experiments}

Experiments were conducted in a glass tank that was 61.5\,cm wide in the horizontal directions and 40\,cm high.  
A section of Perspex, almost as wide as the tank, was attached to the base of the tank, approximately 1\,cm from one wall, via a hinge. 
The Perspex could be rotated to represent a vertical or sloping ice face.

A point source was installed at the centre of the hinged wall and 10\,cm above the base of the tank. 
The source had a radius of 0.27\,cm and was designed such that the discharge was turbulent from the point of release, as described in \cite{Kaye04}. 
The source rotated with the hinged wall such that the discharge was always parallel to the wall and upwards. 
The fountain typically became attached to the wall after a few centimetres.

The tank was initially filled with a mixture of oceanic salt water and fresh water to provide a predetermined density. 
The density was measured using an Anton Parr densimeter to an accuracy of $10^{-6}~\rm{g\,cm^{-3}}$. 
The temperature of the ambient fluid was thermally equilibrated at room temperature by resting the fluid in a storage drum for at least 12\,hours prior to filling the tank. 
Residual motions caused by filling the experimental tank were left to decay for at least 30\,minutes before the experiment was started.

Negatively buoyant seawater, with a small amount of rhodamine dye added for visualisation, was discharged from the source with a flow rate that was controlled by a pump. 
The flow rate was sufficiently high to impart enough vertical momentum for the negatively buoyant fluid to reach the surface.
Similarly to the ambient fluid, the density was measured prior to an experiment with an Anton Parr densimeter and the fluid was allowed at least 12\,hours to thermally equilibrate. 
The pump could provide flow rates from 2.5--7.5\,$\rm{cm^3\,s^{-1}}$. 
Lower flow rates would have been possible but were avoided to ensure that the discharge was turbulent. 

For most of the experiments (\S\ref{sec:Surface}), the flow was illuminated with a horizontal  light sheet located near the free surface of the tank.
Adjacent to the tank, green LEDs produced light that was passed through a cylindrical lens to form a horizontal sheet of green light with a thickness of approximately 0.5--1\,$\rm{cm}$ in the region of interest.
Despite the lens, the light sheet spread slightly in the vertical direction causing its thickness to slightly increase away from the wall and its intensity to slightly decrease within the upper 0.5--1\,$\rm{cm}$.
We expect the negatively buoyant jet to be visible near the free surface until its upper surface falls below the base of the light sheet (i.e. 0.5--1\,cm below the free surface).
For experiments that were designed to measure the horizontal spreading rate of the fountain (\S\ref{sec:Spreading}) the light sheet was placed horizontally at varying depths in the water column. 
For these experiments the light sheet had a thickness of approximately $0.5\,\rm{cm}$ in the region occupied by the fountain.

The fountain surface expression was recorded with a Nikon camera placed directly above the tank. 
The camera recorded a video of the entire experiment that was later processed using `Streams' \citep{Nokes14}. 
Approximately 30\,s of video was time averaged to remove the turbulent fluctuations in the surface expression. 
Letting the experiment continue for longer times resulted in the sinking surface expression fluid entraining back into the fountain. 
Such a situation would not be expected in a geophysical context primarily because the surface expression is considerably larger than the fountain, but also due to further mixing and advection within the fjord that was not present in the laboratory experiments.
A reference image from before the fountain was started was subtracted from the averaged experimental image to remove any effects from inconsistent lighting. 
Finally the intensity of red light was calculated for each pixel of the averaged and subtracted image and a threshold was applied to the resulting intensity field to determine the edge of the dyed fluid (e.g. figure~\ref{fig:plume_subsurface}).

We note that the dye intensity shouldn't be interpreted as a quantitative measure of the dye concentration throughout the surface expression.
Firstly, the dye concentration is uncalibrated and there is no reason to expect a linear relationship between concentration and the intensity of reflected light observed by the camera.
Secondly, across the length of the surface expression we would expect the light sheet to be significantly attenuated.
Having said this, the results do suggest that the observed light intensity primarily decreases due to the advection of dyed fluid below the light sheet.
Several experiments were conducted with the entire tank lit by ambient lighting which allowed the sinking of dye from the free surface to the base of the tank to be qualitatively observed.
Furthermore, as discussed later in \S\ref{sec:Surface}, the measured surface expression dimensions were relatively insensitive to the chosen intensity threshold.
This insensitivity is consistent with advective sinking of dye but would not be expected if attenuation of light or mixing of the dyed fluid was causing the observed reduction in intensity.

\section{Fountain spreading rate}
\label{sec:Spreading}

A small set of experiments was conducted to measure the rate at which the rising fountain spreads horizontally due to entrainment of ambient fluid (i.e. the spreading in region 1 of the theoretical model). 
From \citet{Ezhova18} we expect the flow to spread more rapidly in the wall-parallel direction than the wall-perpendicular direction. 
As such, these experiments had two purposes. 
First, to measure the relative length of the fountain in the wall-parallel and the wall-perpendicular directions and second, to measure a bulk entrainment coefficient for use in the fountain model described in \S\ref{sec:Theory}\ref{sec:phase1}.

We assume that the wall fountain is a semi-ellipse and calculate the top-hat radius of an equivalent semi-circular fountain with the same cross-sectional area as
\begin{equation}
  b = \sqrt{r_{1} r_2},
  \label{eq:effective_radius}
\end{equation}
where $r_1$ and $r_2$ are the measured half-widths of the rising fountain in the wall-parallel and wall-perpendicular directions, respectively.

\begin{figure}
\centering
  \noindent\includegraphics[width=19pc]{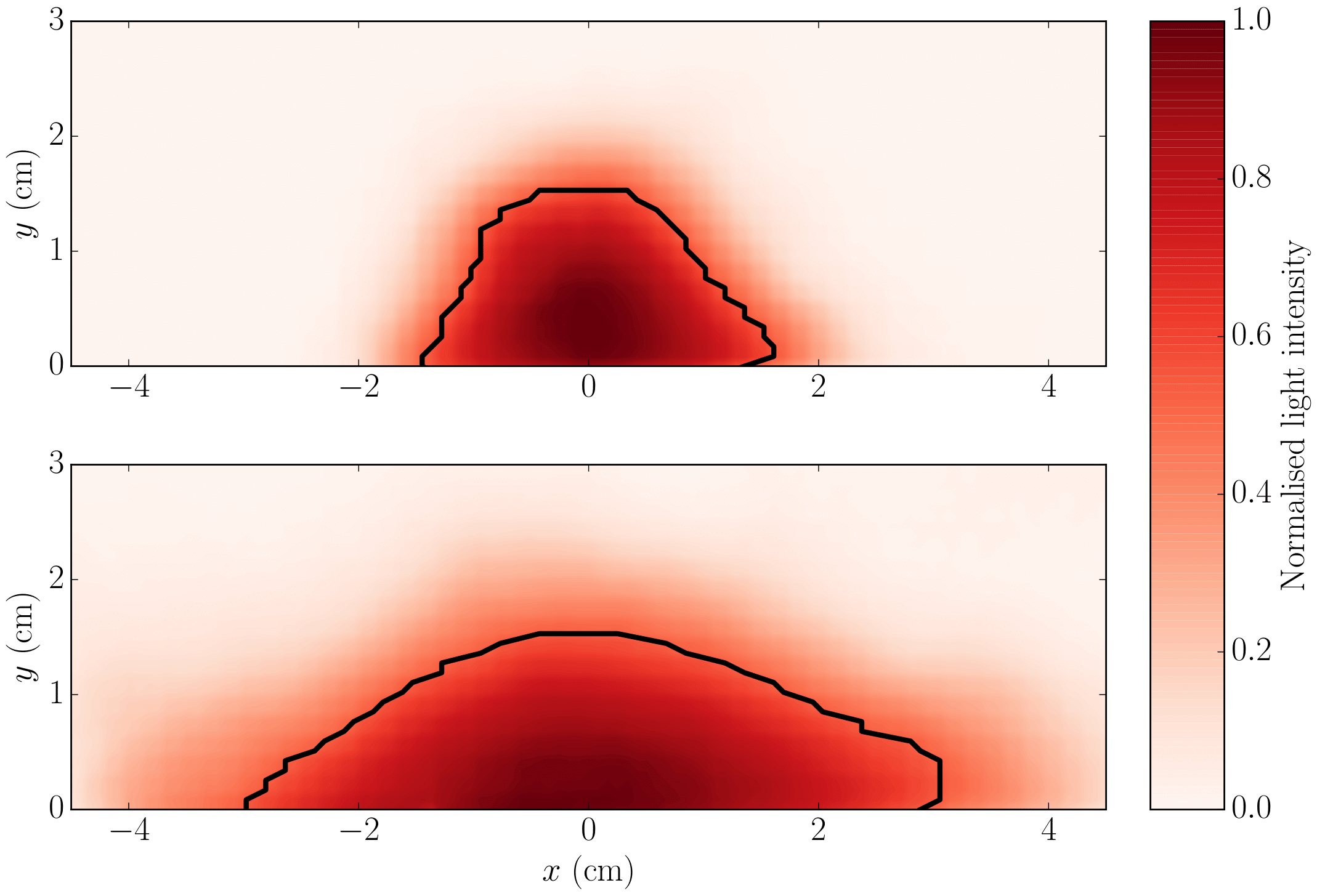}
  \caption{Normalised light intensity showing the horizontal subsurface spreading of the fountain.
  The top image shows the fountain 4.8\,cm above the source while the bottom image shows the fountain 10.5\,cm above the source. 
  The black line shows the 52\% contour level used to determine the top-hat fountain width.}
  \label{fig:plume_subsurface}
\end{figure}

Figure~\ref{fig:plume_subsurface} shows two images from the spreading rate experiments.
The images have been processed as described in \S\ref{sec:Experiments} to show the normalised light intensity. 
The top image shows the fountain shape at a height of 4.8\,cm above the source while the second image shows the fountain 10.5\,cm above the source. 
It is clear that the fluid has spread much more rapidly in the wall-parallel direction than in the wall-perpendicular direction, leading to an increasing asymmetry with height.
The black lines on figure~\ref{fig:plume_subsurface} show the fountain edge based on a threshold intensity of 52\% of the maximum value. 
A threshold value of 52\% is used as representative of a top-hat profile where the concentration field spreads slightly more rapidly than the velocity field \citep{Turner73}.

\begin{figure}
\centering
  \noindent\includegraphics[width=19pc]{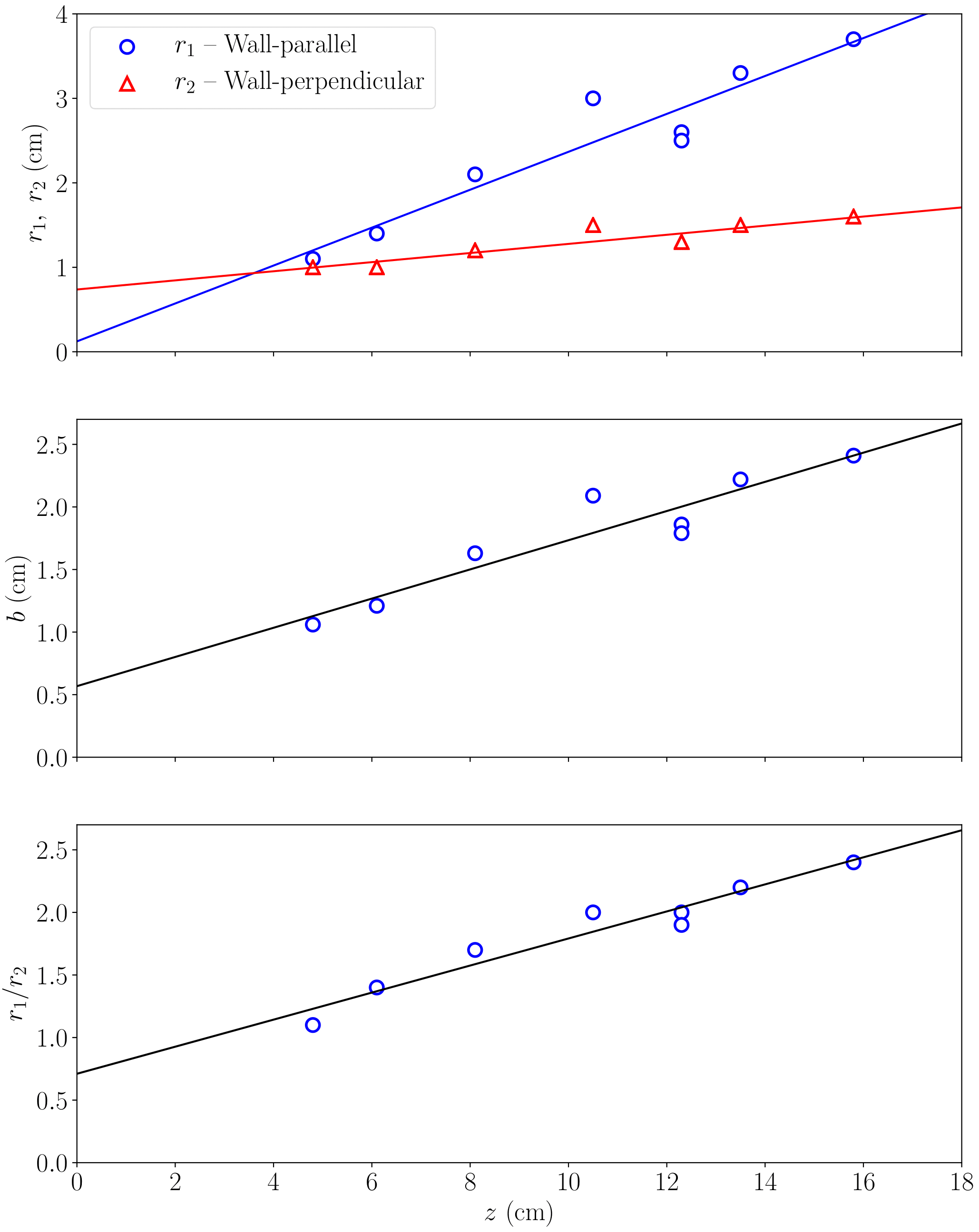}
  \caption{Experimental measurements of horizontal fountain spreading as a function of height above the source. 
  The top panel shows the measured half widths in the wall-parallel and \mbox{wall-perpendicular} directions, the middle panel shows the top-hat radius of an equivalent semi-circular fountain and the bottom panel shows the ratio of half widths in the wall-parallel ($r_1$) to the wall-perpendicular ($r_2$) direction.}
\label{fig:plume_spreading}
\end{figure}

Figure~\ref{fig:plume_spreading} shows all measurements of the top-hat fountain half-widths in the wall-parallel and wall-perpendicular directions as a function of height above the source. 
Also shown is the equivalent radius of a semi-circular fountain from equation~(\ref{eq:effective_radius}). 
A linear regression has been used to calculate the spreading rate of the fountain which is used to calculate an entrainment coefficient based on the canonical self-similar plume model \citep{Morton56}:
\begin{equation}
  \alpha = \frac{5}{6}\,\frac{{\rm{d}}b}{{\rm{d}}z}.
\end{equation}

For the range of heights that were tested, the semi-circular fountain radius is given by
\begin{equation}
  b = 0.12z+0.57
\end{equation}
where $b$ and $z$ are measured in cm and $z$ is measured from the source. 
Thus, the bulk entrainment coefficient that we used in \S\ref{sec:Theory}\ref{sec:phase1} is $\alpha=\,0.10$, consistent with the range of values typically found for turbulent jets and plumes \citep{Carazzo16}. 
The ratio of half-widths in the wall-parallel and wall-perpendicular directions is given by
\begin{equation}
  \frac{r_1}{r_2} = 0.11 z+0.71.
\end{equation}
This ratio, as well as the calculated fountain radius from (\ref{MTT_Q})--(\ref{MTT_B}), is used to calculate $m$ and $n$ in (\ref{eq:wxy}) and (\ref{eq:gxy}). 
Since our source is circular we would expect the initial ratio of half-widths to be 1. 
The lower value of 0.71 is likely due to the fountain being drawn towards the wall by the Coand\u{a} effect \citep{Wille65} whereby entraining flows create a low pressure region near boundaries and hence are attracted to the boundary.

We note that far away from the source it is expected that the aspect ratio would reach a constant value given by the ratio of the spreading rate in the wall-parallel direction to that in the wall-perpendicular direction. 
Thus, although the aspect ratio is seen to increase with height for the experimental fluid depths used in the present study, we would expect it to be constant at the greater fluid depths relevant to geophysical situations. 
We estimate this constant aspect ratio from the upper panel of figure~\ref{fig:plume_spreading} as
\begin{equation}
\left.\left(\frac{{\rm{d}}r_1}{{\rm{d}}z}\right) \middle/ \left(\frac{{\rm{d}}r_2}{{\rm{d}}z}\right)\right. \approx 4.
\label{eq:AR}
\end{equation}
We note that the ratio of spreading rates is very similar to that from numerical simulations of a buoyant plume next to a vertical wall \citep{Ezhova18}.

\section{Surface expression}
\label{sec:Surface}

The majority of the experiments were designed to examine the surface expression of the fountain. 
The initial volume flux and reduced gravity of the fountain, as well as the ambient fluid depth, were all varied over the experiments. 
These changes are equivalent to changing the properties of the plume penetrating into the upper layer of the fjord and changing the depth of the upper layer.

\cite{Burridge17} considered a two-dimensional dense jet released horizontally at the free surface and found that the non-dimensionalised distance that the jet stays at the surface increases linearly with the source Froude number defined as ${\rm{Fr}}={u}{(g'\delta)^{-1/2}}$. 
Although the flow that we are considering is significantly different from a two-dimensional surface jet, we expect that the length and width of the surface expression in our experiments will have a similar dependence on Fr.

Values of the experimental parameters and calculated values of $\delta$ and $\rm{Fr}$ are provided for each of the experiments in table~\ref{tab:experiments}. 
The values of $\delta$, $u$, and $g'$ (and hence Fr) are calculated at the transition to the horizontal negatively buoyant jet region based on the model presented in \S\ref{sec:Theory}. 
Since the values of $\delta$ and $\rm{Fr}$ are different for each sector in the model we have shown the two extreme values: 
the wall-parallel direction $x$, and the wall-perpendicular direction $y$.

\begin{table}
  \centering
  \begin{tabular}{lcccccc}
    $Q_0$	& $g'_0$ & $H$ &$\delta_x$ &$\delta_y$&${\rm{Fr}}_x$ & ${\rm{Fr}}_y$\\
    ($\rm{ml\,s}^{-1}$)& ($\rm{cm\,s}^{-2}$) & (cm)  &(cm) &(cm)&(-)&(-) \\[3pt]
    2.71	& 3.54	& 10.0	& 1.23	& 1.52	& 1.17	& 1.44 \\ 
    5.89	& 12.10	& 11.2	& 1.24	& 1.50	& 1.43	& 1.81 \\ 
    4.30	& 7.00	& 10.0	& 1.08	& 1.33	& 1.69	& 2.08 \\ 
    5.89	& 10.61	& 11.0	& 1.10	& 1.38	& 1.82	& 2.25 \\ 
    3.42	& 3.32	& 10.0	& 0.98	& 1.22	& 2.11	& 2.86 \\ 
    5.89	& 9.18	& 11.0	& 1.05	& 1.31	& 2.13	& 2.66 \\ 
    4.30	& 3.73	& 10.0	& 0.96	& 1.19	& 2.59	& 3.09 \\ 
    4.71	& 2.80	& 13.1	& 1.14	& 1.32	& 3.05	& 4.08 \\ 
    5.09	& 4.06	& 10.0	& 0.89	& 1.10	& 3.83	& 4.72 \\ 
    5.89	& 4.57	& 11.0	& 0.93	& 1.16	& 3.87	& 4.82 \\ 
    5.89	& 4.44	& 10.0	& 0.88	& 1.08	& 4.36	& 5.42 \\ 
    5.89	& 3.32	& 10.0	& 0.86	& 1.06	& 5.28	& 6.52 \\ 
    7.48	& 4.56	& 10.0	& 0.85	& 1.06	& 5.85	& 7.13 \\ 
    6.39	& 3.61	& 9.0	& 0.78	& 0.99	& 6.23	& 7.44 \\ 
    6.68	& 2.92	& 10.0	& 0.85	& 1.05	& 6.60	& 8.13 \\ 
    6.72	& 3.68	& 8.5	& 0.67	& 0.95	& 7.28	& 8.18 \\ 
  \end{tabular}
  \caption[Experimental properties]
  {Source volume flux $Q_0$, source reduced gravity $g'_0$, free surface height above the source $H$, and calculated values of $\delta$ and Fr for all of the surface expression experiments with a vertical wall. 
  $Q_0$, $g'_0$, and $H$ are external parameters that are measured based on the source and ambient properties while $\delta$ and Fr are calculated at the transition to the negatively buoyant jet based on the model presented in \S\ref{sec:Theory}.}
  \label{tab:experiments}
\end{table}

\subsection{Dimensions of the surface expression}

\begin{figure}
\centering
  \noindent\includegraphics[width=19pc]{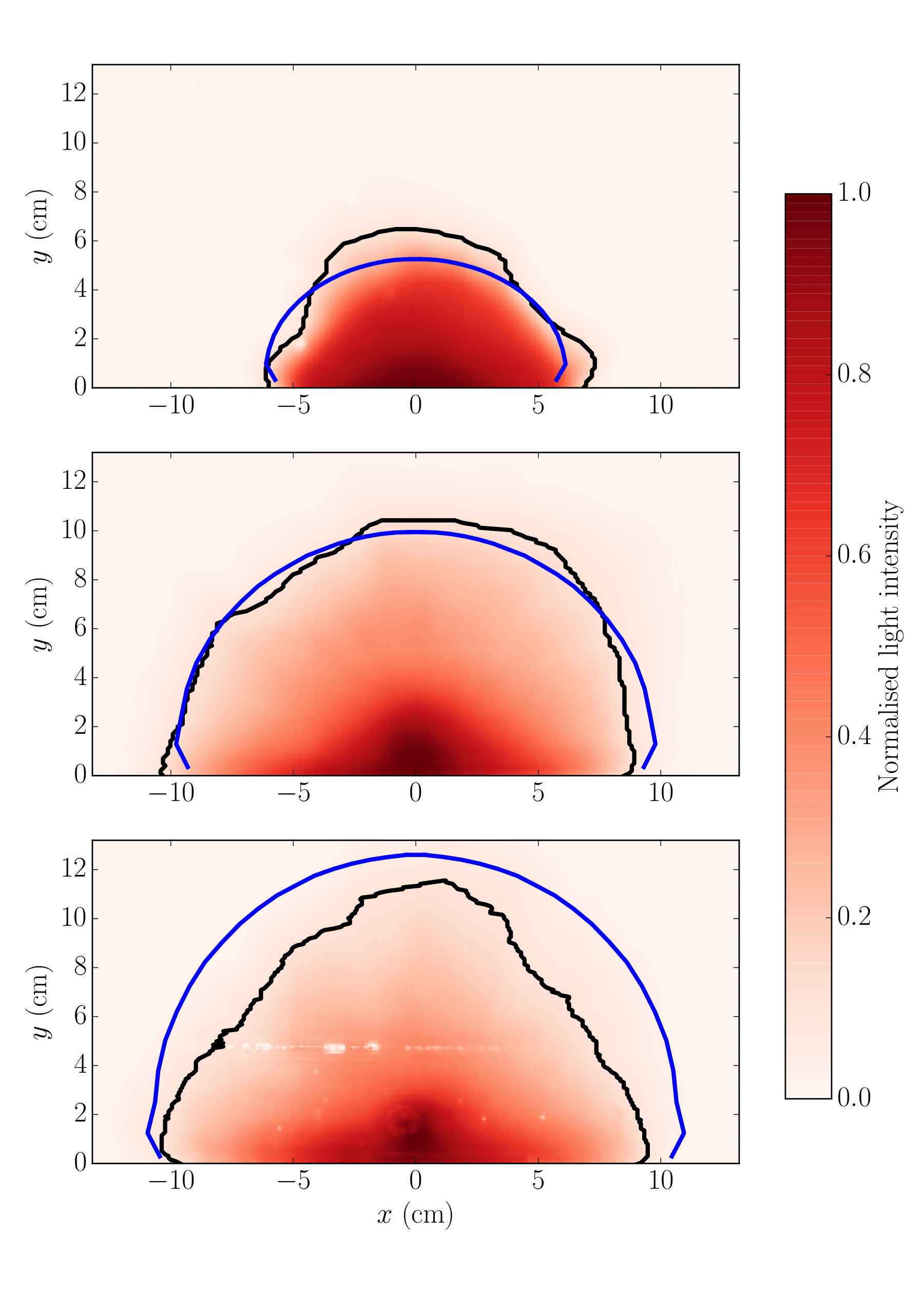}
  \caption{Normalised light intensity as measured from typical experiments focused on the surface expression. 
  The particular experiments are those with ${\rm{Fr}}_y =1.44$ (top),~4.82 (middle), and~8.18 (bottom) as given in table~\ref{tab:experiments}. 
  The black line shows the normalised 10\% light intensity threshold which was used to determine the edge of the surface expression and the blue line shows the predicted shape of the surface expression based on the model presented in \S\ref{sec:Theory}.}
  \label{fig:plume_surface}
\end{figure}

Figure~\ref{fig:plume_surface} shows processed images of the surface expression for three experiments. 
The edge of the surface expression, defined by the normalised 10\% light intensity contour, is shown in black and the model prediction of the surface expression given by $(s_x,s_y)$ from (\ref{sx})--(\ref{sz}) is superimposed in blue. 
A 10\% threshold was used, different from the 52\% threshold used to determine the top-hat fountain width (figure~\ref{fig:plume_spreading}), because we are interested in the actual size of the surface expression rather than a top-hat scale.
As such, we selected the minimum threshold possible while remaining above the noise levels of the experimental observations.
The model prediction shows reasonable agreement with the experiments but for large values of Fr tends to predict a more semi-circular surface expression than is observed in the experiments.

The reduction in observed light intensity is caused by a variety of processes and is not correlated directly with the concentration of dye within the surface expression. 
The relatively low sensitivity of the observed surface expression dimensions to changes in the intensity threshold defining the edge of the surface expression (described later) suggests that the primary process reducing the observed light intensity is the advective sinking of dyed fluid below the light sheet. 
However, there are a number of secondary processes that could also cause the observed reduction of the light intensity. 
The most important of these are the attenuation of the light sheet as it passes through the dyed fluid, averaging the temporal variability when producing the light intensity figures, and dilution of the surface expression dyed fluid due to mixing. 
Dilution of the surface expression is increasingly important for experiments with larger Froude numbers. 
Since the experiments were not designed to measure dye concentration (which would have required a camera with a larger dynamic range and careful calibration) we are unable to quantify the effects of dilution in the experiments. 
However, based on our model of the surface expression flow, we expect the dye to dilute to 85\% and 43\% of its maximum value across the surface expression for the top and bottom panel of figure~\ref{fig:plume_surface}, respectively. 
At high Fr numbers both the dilution and the attenuation of light are significant which could help to explain why the observed surface expression is smaller than the model predictions for large Fr.

\begin{figure}
\centering
  	\noindent\includegraphics[width=19pc]{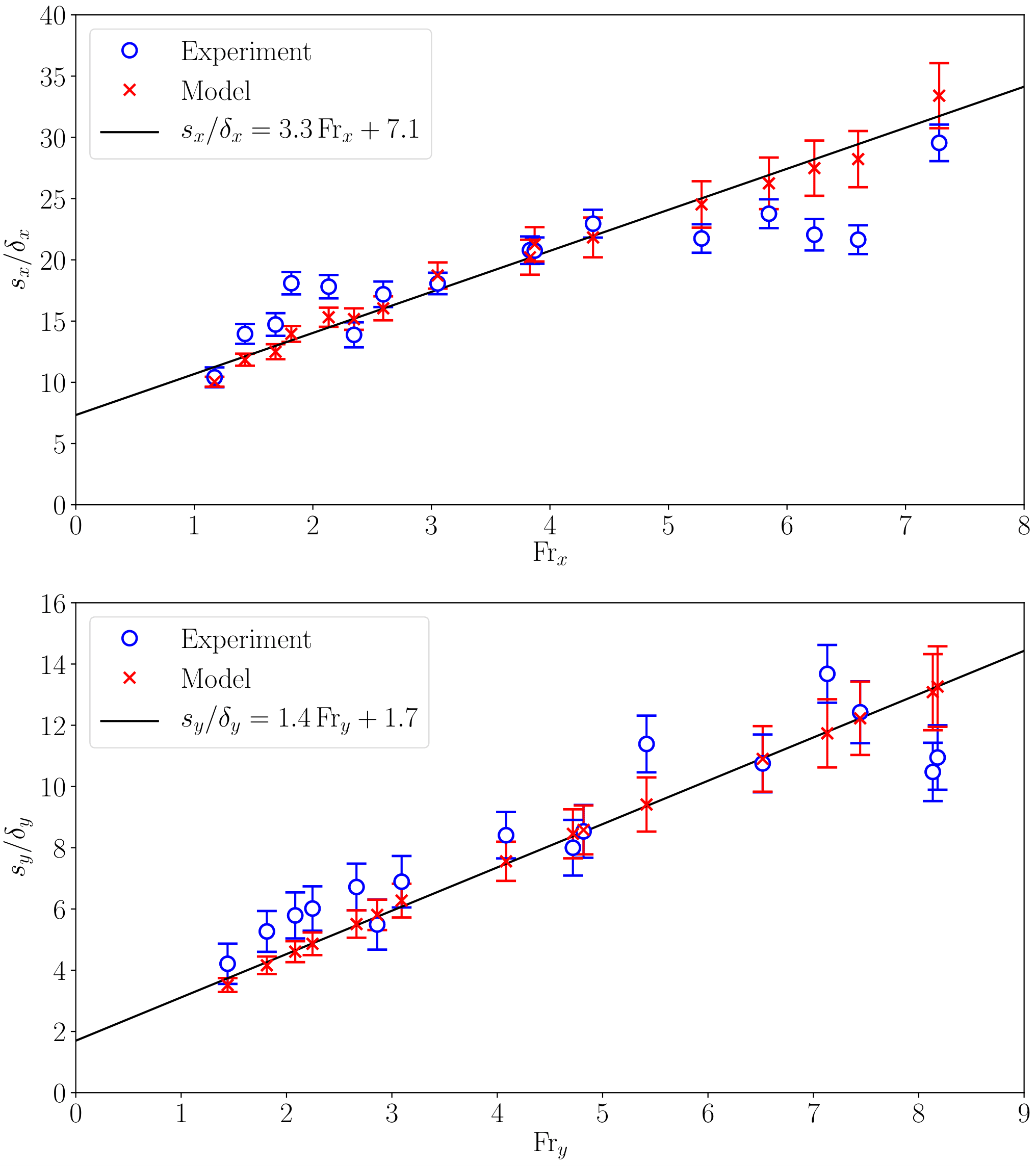}
  	\caption{Experimental measurements of the fountain surface expression half-width ($s_x$, top panel) and length ($s_y$, lower panel) plotted as a	function of ${\rm{Fr}}_x$ and ${\rm{Fr}}_y$, respectively. 
  	The length and half-width of the surface expression are non-dimensionalised by the jet	thickness $\delta$ in the corresponding direction. 
  	Also shown are the model predictions for each experiment with a linear fit to the model predictions.}
  	\label{fig:surface_expression}
\end{figure}

Figure~\ref{fig:surface_expression} shows the experimental and predicted length and half-width of the surface expression, non-dimensionalised by $\delta$, for each experiment given in table~\ref{tab:experiments}.
The error bars for the model predictions are calculated by using a depth threshold for the negatively buoyant jet of 0.5\,cm and 1\,cm to reflect the uncertain position of the bottom of the light sheet during the experiments. 
The error bars for the experimental results are estimated to be 1\,cm for all experiments.
This is based on processing the experimental data with intensity thresholds of 5\% and 20\% when defining the edge of the surface expression for a selection of experiments. 
The relatively low sensitivity of the measured surface expression dimensions to the chosen intensity threshold suggests that the light intensity is decreasing due to the dyed surface expression fluid sinking below the light sheet rather than due to mixing or attenuation of the light sheet. 
Once the surface expression fluid starts sinking, the light intensity will quickly decrease as the dyed fluid sinks below the thin light sheet. 
In contrast, both mixing and light attenuation will reduce the observed light intensity continually with a initially rapid decrease followed by a slower decay.
  
A linear fit is plotted through the model predictions shown in figure~\ref{fig:surface_expression}. 
The linear fit is applied to the model predictions rather than the experimental values to test the similarity of our negatively buoyant jet model to that of \cite{Burridge17}. 
The applicability of the model presented by \cite{Burridge17} is not obvious \textit{a priori}. 
\cite{Burridge17} give the applicability of their model to flows where the Froude number
is greater than 12 whereas the Froude numbers in the present study vary between 1 and 9.
Additionally, the flows being considered are significantly different. 
\cite{Burridge17} considered a two-dimensional jet that only spreads vertically due to entrainment from the base, while we consider a jet that also spreads azimuthally as it propagates. 
However, figure~\ref{fig:surface_expression} shows that the model presented in \S \ref{sec:Theory} also results in the non-dimensionalised surface expression being linearly dependent on ${\rm{Fr}}$ (with a different constant of proportionality compared to the results of \cite{Burridge17}).
The dimensions of the surface expression are accurately predicted by the model presented in \S\ref{sec:Theory} across most of the parameter space with a root mean square error of 2.95  and 1.30 for $s_x/\delta_x$ and $s_y/\delta_y$, respectively (approximately 15\% in each case).

\subsection{Aspect ratio of the surface expression}

\begin{figure}
\centering
  \noindent\includegraphics[width=19pc]{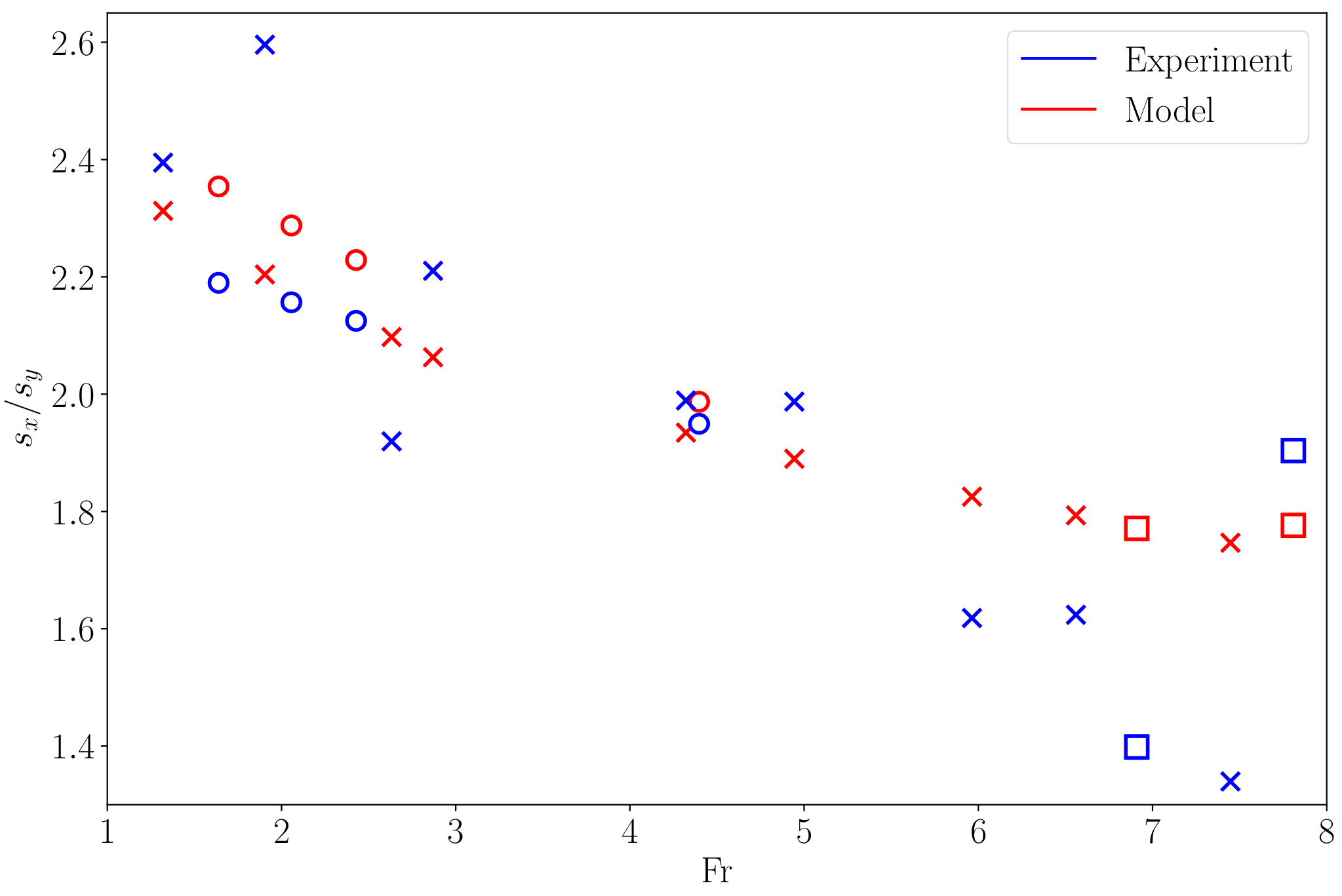}
  \caption{Experimental and predicted values of the surface expression aspect ratio as a function of ${\rm{Fr}}$. 
  The value of ${\rm{Fr}}$ used is an average of ${\rm{Fr}_x}$ and ${\rm{Fr}_y}$.
  Experimental results are shown in blue while model predictions are shown in red.
  Experiments with a fluid depth of 10\,cm are shown by crosses, a fluid depth of 11--11.2\,cm by circles, and a fluid depth of 8.5--9\,cm by squares.}
  \label{fig:aspect}
\end{figure}

We define an aspect ratio of the surface expression as the width parallel to the wall ($s_x$) divided by the length perpendicular to the wall ($s_y$). 
Therefore, if the surface expression was semi-circular it would have an aspect ratio of 2.
Figure~\ref{fig:aspect} shows the experimental and predicted aspect ratio of the surface expression for the experiments given in table~\ref{tab:experiments}. 
The different symbols indicate different fluid depths. 
The predicted values are shown in red while the experimental values are shown in blue. 
The aspect ratio is plotted against ${\rm{Fr}}$ which is the average of ${\rm{Fr}}_x$ and ${\rm{Fr}}_y$.

Figure~\ref{fig:aspect} shows that the aspect ratio decreases as ${\rm{Fr}}$ increases.
For low values of Fr the surface expression is predominantly defined by the shape of the fountain below the surface (figure~\ref{fig:plume_subsurface}, bottom panel) and the aspect ratio is larger than 2. 
As Fr increases, the negatively buoyant jet travels further away from the wall before sinking below the light sheet and the initial asymmetry in fountain shape becomes less important. 
Instead, the lower radius of curvature at the middle of the fountain and the offset velocity and density profiles lead to the jet travelling further away from the wall than along the wall (\S\ref{sec:Theory}\ref{sec:phase3}) and the aspect ratio decreases.

\subsection{The effect of a sloping wall}
\label{sec:slope}

A supplementary set of experiments was undertaken to investigate the effect of a sloping wall on the surface expression. 
Experiments similar to those described in \S\ref{sec:Spreading} showed that the entrainment coefficient and fountain asymmetry were not significantly affected by a sloping wall. 
Due to refraction of light from the Perspex wall, visualising the fountain beneath the surface was much more challenging for a sloping wall than for a vertical wall. 
As such, the results were not used to determine the entrainment coefficient, only to confirm that the value is not significantly different from the vertical case. 
Furthermore, we assume that the distance that the maximum velocity is offset from the wall is unaffected by the slope.

The model that was presented in \S\ref{sec:Theory} is slightly adapted to account for a sloping wall. 
In the vertical wall case, all of the fountain momentum that entered into a sector was converted to horizontal momentum with a direction normal to the sector boundary.
For a sloping wall, the same process is undertaken for the vertical component of the fountain momentum, but the horizontal momentum is retained in the wall-perpendicular direction.
This adjustment results in the length of the surface expression ($ s_y $) being increased and the width of the surface expression ($ s_x $) being decreased. 
The effect on the length is small as the sector boundary in the centre of the surface expression is approximately parallel to the wall. 
As such, both the vertical and horizontal components of the fountain momentum leave the sector in the wall-perpendicular direction ($\beta=90^{\circ}$ on figure~\ref{fig:Surface_deviation}). 
In contrast, for the sector next to the wall, the vertical component of the fountain momentum will be directed along the wall while the horizontal component will be directed in the wall-perpendicular direction. 
More significant changes to the length would require transfer of momentum between sectors and it is not clear how this should be done.

\begin{figure}
\centering
\noindent\includegraphics[width=19pc]{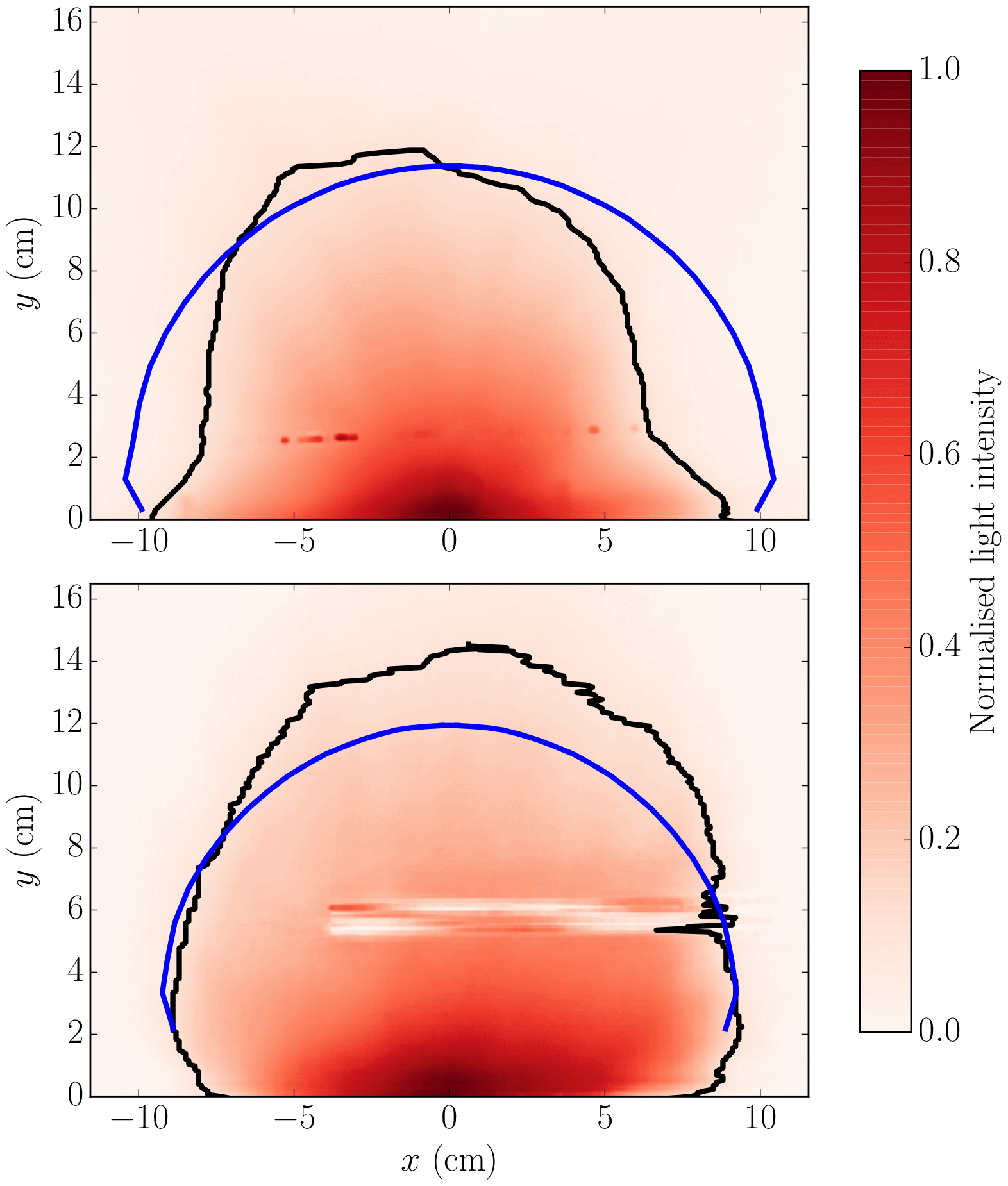}
\caption{Normalised light intensity as measured from two typical experiments. 
The upper panel shows an experiment with a vertical wall and $Q_0 = 5.89\,\rm{ml\,s^{-1}}$,  $g'_0= 3.32\,\rm{cm\,s^{-2}}$ while the lower panel shows an experiment with $\theta = 55^{\circ}$ and $Q_0=5.89\,\rm{ml\,s^{-1}}$, $g'_0= 3.52\,\rm{cm\,s^{-2}}$. 
The low light intensity around $y=5\,{\rm{cm}}$ on the lower panel is an artefact of the top of the Perspex wall and is not physically meaningful.
The blue line is showing the model prediction and the black line is showing the normalised 10\% intensity threshold from the experiments.}
\label{fig:surface_slope}
\end{figure}

The experimental parameters for the sloping experiments are given in
table~\ref{tab:slope}. 
Figure~\ref{fig:surface_slope} shows the normalized light intensity field for two experiments with similar discharge characteristics but a vertical and a sloping wall. 
It can be seen that in the case of a sloping wall, the fountain fluid travels less distance in the wall-parallel direction and a greater proportion of the fluid ends up far away from the wall.

\begin{table}
\centering
\begin{tabular}{cccccccc}
 $\theta$ & $Q_0$	& $g'_0$			 & $H$ 		&$\delta_x$ & $\delta_y$	& ${\rm{Fr}}_x$	 & ${\rm{Fr}}_y$	\\
(${}^\circ$) & ($\rm{ml\,s}^{-1}$) 	& ($\rm{cm\,s}^{-2}$) & (cm)  &(cm) &(cm) & (-) & (-) \\[3pt]
55	& 3.63	& 4.25	& 10.0	& 1.52	& 1.34	& 1.14	& 2.03 \\ 
55	& 4.30	& 4.32	& 10.0	& 1.38	& 1.22	& 1.62	& 3.26 \\ 
55  & 5.13	& 3.52	& 10.0	& 1.27	& 1.11	& 2.55	& 5.20	\\ 
55	& 5.89	& 4.25	& 10.0	& 1.26	& 1.11	& 2.71	& 5.47 \\ 
55  & 5.89 	& 3.52	& 10.0	& 1.24	& 1.09	& 3.07	& 6.23	\\ 
55	& 5.89	& 4.98	& 10.0	& 1.23	& 1.08	& 3.26	& 6.64 \\ 
55	& 7.48	& 4.32	& 10.0	& 1.22	& 1.07	& 3.65	& 7.41	\\ 

70	& 3.54	& 4.25	& 10.0	& 1.27	& 1.24	& 1.43	& 2.50	\\ 
70  & 4.30 	& 3.52	& 10.0	& 1.12 	& 1.09	& 2.46	& 4.32	\\ 
70  & 5.13 	& 3.52	& 10.0	& 1.07	& 1.05	& 3.21	& 5.60	\\ 
70	& 5.89	& 4.57	& 10.0	& 1.07	& 1.04	& 3.24	& 5.69	\\ 
70	& 7.48	& 4.25	& 10.0	& 1.03	& 1.01	& 4.61	& 8.03	\\ 
\end{tabular}
\caption[Experimental properties]
{Experimental slope angle $\theta$, source volume flux $Q_0$, source reduced gravity $g'_0$, free surface height above the source $H$, and calculated values of $\delta$ and ${\rm{Fr}}$ for all of the surface expression experiments with a sloping wall. 
The angle $\theta$ is measured from the horizontal.}
  \label{tab:slope}
\end{table}

\begin{figure*}
\centering
\noindent\includegraphics[width=27pc]{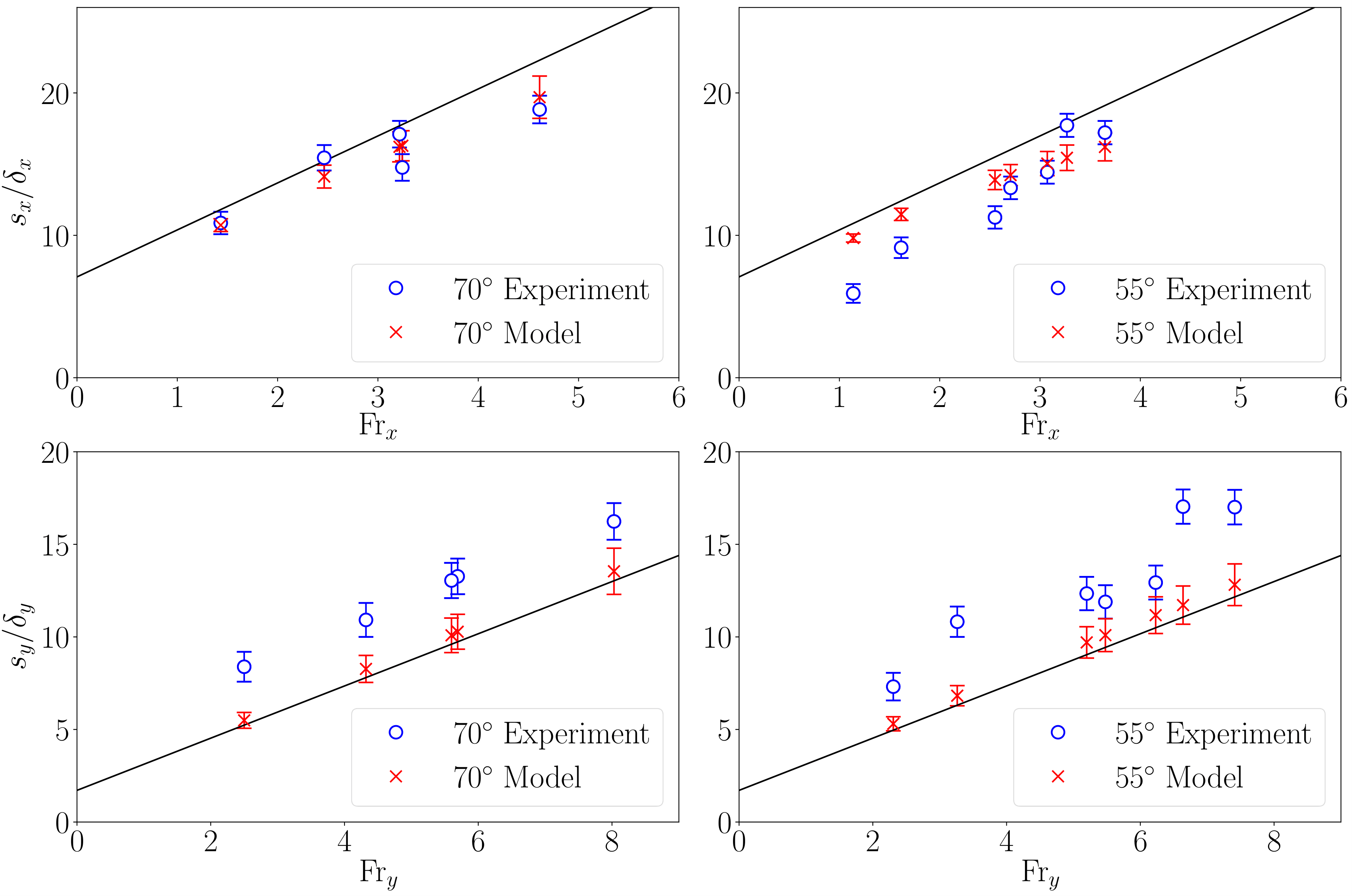}
\caption{Experimental measurements and model predictions for the fountain surface expression for experiments with a sloping wall. 
The top row shows the wall-parallel ($s_x$) direction and the bottom row shows the wall-perpendicular ($s_y$) direction. 
The left column shows experiments with a slope angle of $70^\circ$ and the right column shows experiments with a slope angle of $55^\circ$. 
The solid lines are the same linear fit shown in figure~\ref{fig:surface_expression} for a vertical wall.}
\label{fig:slope}
\end{figure*}

Figure~\ref{fig:slope} gives the experimental and model results for experiments conducted with a sloping wall and shows that the model predictions do not deviate significantly from the linear dependence on ${\rm{Fr}}$ that was determined for a vertical wall (solid lines on figure~\ref{fig:slope}).  
We note that although the model results shown on figure~\ref{fig:slope} suggest that the dependence of the surface expression size on Fr is not significantly affected by the presence of a sloping wall, the modifications to the theoretical model described at the beginning of this section can have a large impact on the value of Fr. 
As an example, the experiment shown in table~\ref{tab:experiments} for a vertical wall with $Q_0=5.89\,\rm{ml\,s^{-1}}$ and $g'_0 = 4.57\,\rm{cm\,s^{-2}}$ has Froude number values of ${\rm{Fr}}_x=3.87$ and ${\rm{Fr}}_y=4.82$. 
In contrast, a similar experiment with a $55^{\circ}$ slope angle ($Q_0=5.89\,\rm{ml\,s^{-1}}$ and $g'_0 = 4.98\,\rm{cm\,s^{-2}}$ on table~\ref{tab:slope}) has Froude number values of ${\rm{Fr}}_x=3.26$ and ${\rm{Fr}}_y=6.64$.
However, even with the modified treatment of the fountain momentum through the transition region, the model underpredicts the length of the surface expression for both $55^\circ$ and $70^\circ$ angles.

\begin{figure}
\centering
\noindent\includegraphics[width=19pc]{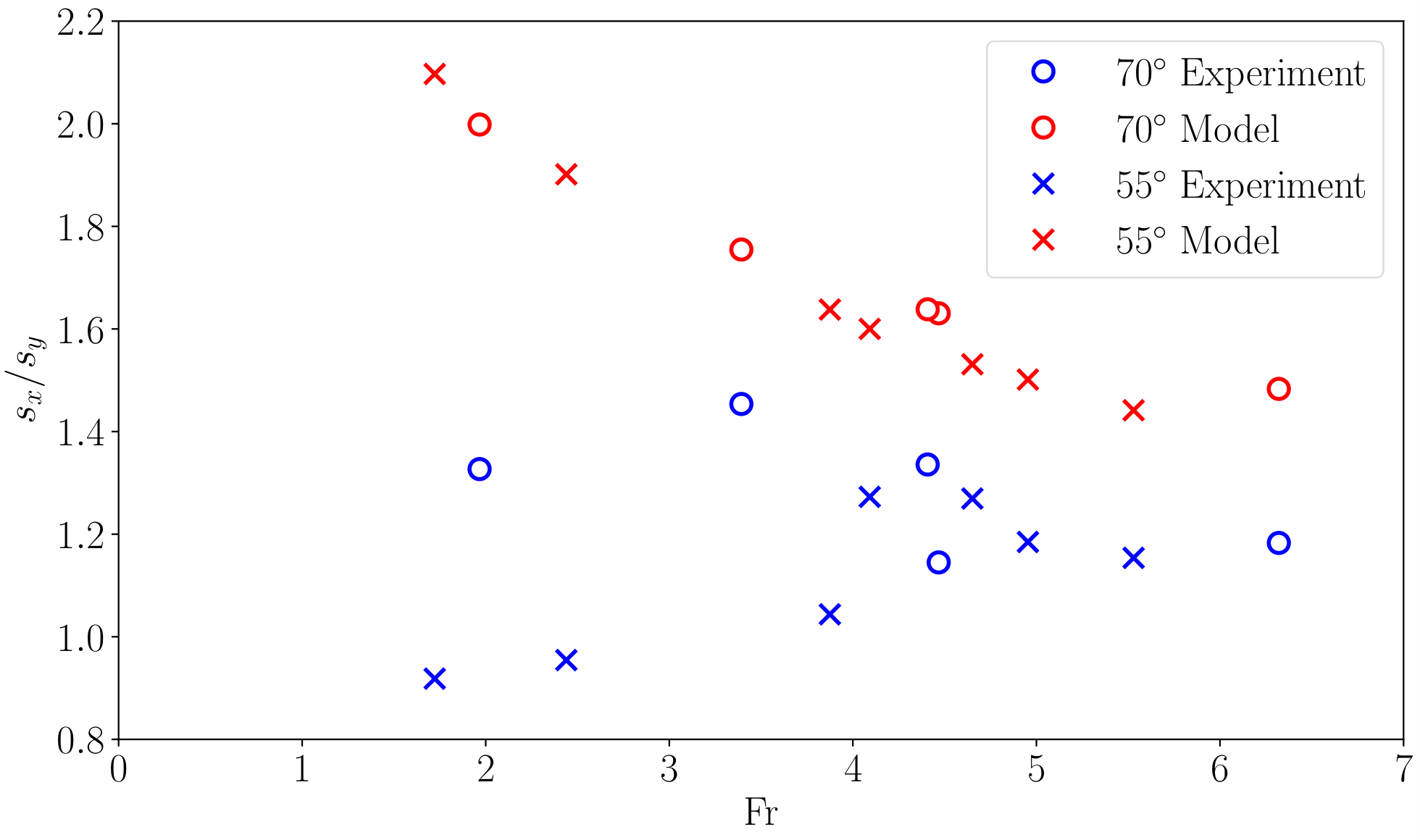}
\caption{Experimental and predicted values of the surface expression aspect ratio as a function of Fr for the experiments with a sloping wall.
The value of Fr used is the average of $\rm Fr_x$ and $\rm Fr_y$.}
\label{fig:slope_ar}
\end{figure}

Figure~\ref{fig:slope_ar} shows the measured and predicted values of the surface expression aspect ratio for the sloping wall experiments. 
Similarly to figure~\ref{fig:aspect}, the value of Fr that is shown is the mean value of ${\rm{Fr}}_x$ and ${\rm{Fr}}_y$. 
Since the model systematically under predicts the length of the surface expression, the predicted aspect ratio is always too large. 
The discrepancy between the predicted and measured aspect ratio is larger for the $55^{\circ}$ experiments than for the $70^{\circ}$ experiments as the underestimation of the surface expression length is larger. 
The discrepancy is reduced for larger values of Fr as the surface expression becomes larger and the effect of an approximately constant error in the wall-perpendicular direction is reduced.

\section{Application to observations}
\label{sec:Application}

In this section we apply the model presented in \S \ref{sec:Theory} to observations of a subglacial discharge plume from Saqqarliup Fjord, Greenland \citep{Mankoff16}. 
Photographs of the fjord surface show a triangular surface expression that extends approximately 300\,m along the front of the glacier and 300\,m into the fjord \citep[see figure 3a of][]{Mankoff16}.
Accompanying these aerial photographs of the surface expression are observations of the water properties within the fjord --- both some distance downfjord and through the surface expression. 
The oceanographic observations show a significantly different density profile downstream than through the surface expression.

Downstream in the fjord, the 150\,m water column has an approximately two-layer density profile with warmer and fresher water overlying cooler and saltier water \citep[see figure 7 of][]{Mankoff16}.
We characterise the overlying layer as $S=29.5\,{\rm g\,kg^{-1}}$, $T = 2.0{\rm ^\circ C}$, $\rho = 1023.4\,{\rm kg\,m^{-3}}$ and the underlying layer as $S=33\,{\rm g\,kg^{-1}}$, $T = 1.0{\rm ^\circ C}$, $\rho = 1026.3\,{\rm kg\,m^{-3}}$, where $S$ and $T$ are representative values of the salinity and temperature for each layer taken from figure 7 of \cite{Mankoff16} and $\rho$ is the density at those conditions calculated based on the thermodynamic equation of seawater \citep{TEOS10}.
Throughout the remainder of this section we will refer to these two layers as the upper and lower layers, respectively.

Salinity profiles through the surface expression show that the density profile is significantly altered with a weak and more uniform stratification \citep[see figure 5 of][]{Mankoff16}.
It is expected that during periods when the subglacial discharge is not present, as is generally assumed throughout winter, the two-layer stratification will exist throughout the entire fjord but that the strong subglacial discharge displaces the upper layer down the fjord.

In addition to the photographic observation of the surface expression and temperature-salinity profiles, \cite{Mankoff16} attempted to infer the subglacial discharge flux based on observed water mass properties.
They estimated the subglacial discharge flux at the base of the glacier to be $105-140\,{\rm m^3\,s^{-1}}$, which compares favorably with estimated runoff for the 5 days prior to observation from the RACMO model of $101\,{\rm m^3\,s^{-1}}$.
In their attempts at modelling the flow, \cite{Mankoff16} assumed that the radius of the subglacial discharge source was $5-15$\,m.

We will apply the model presented in \S\ref{sec:Theory} to the observations in two distinct ways.
The first of these is most analogous to the laboratory experiments and corresponds to the expected conditions when the plume first develops at the start of the melt season.
As mentioned above, without a strong subglacial discharge, we expect the two-layer stratification observed away from the surface expression in \cite{Mankoff16} to exist across the entire fjord.
The fluid immediately below the surface expression is taken to have properties equal to the upper layer.
As such, the plume rises through this two-layer stratification until it hits the free surface and then sinks into the underlying fluid that is less dense than the surface expression.

The second way that we apply the model uses the observations in \cite{Mankoff16} more directly.
Based on the the density profiles through the surface expression \citep[figure 5 of][]{Mankoff16}, two adjustments to the previously described conceptual framework are required.
First, the plume will not entrain fluid from the two-layer stratification which is positioned far downstream but from the fluid that directly surrounds the plume.
The fluid surrounding the plume is best approximated by the salinity and the temperature profiles through the surface expression \citep[dark lines on figure 5 of][]{Mankoff16} --- 
recall that the surface expression is roughly 10 times as large as the expected plume diameter \citep[approximately 20\,m, figure 12a of][]{Mankoff16}, so the profiles through the surface expression are still outside the rising plume.
Using either the downstream density profile or that through the surface expression for entrainment into the vertical plume has a minimal impact on either the vertical plume or the density of the surface expression, but our choice is more consistent with the observations than using the downstream density profile.
The more significant conceptual adaptation that needs to be made arises from the observation that the water column through the surface expression is \emph{stably} stratified in the vertical direction \citep[again, refer to the dark lines on figure 5 of][]{Mankoff16}.
As a result, the surface expression fluid is unable to sink vertically beneath a less dense underlying fluid layer and an alternate mechanism for limiting the surface expression size needs to be considered.

We propose that instead of \emph{sinking} into a less dense underlying layer, the surface expression fluid is \emph{subducted} below the less dense upper layer that is positioned further down the fjord.
A subduction mechanism is reminiscent of an atmospheric cold front where two layers of air with different densities flow into one another and the less dense layer is displaced upwards.
However in this case, instead of less dense air being pushed upwards, the denser fluid layer (the surface expression) is pushed downwards.
The subduction mechanism is consistent with observational data \citep[figure 7 of][]{Mankoff16} which shows strongly sloping isopycnals near the edge of the surface expression but no evidence of dense fluid on top of less dense fluid, as is required by the sinking mechanism explored in the experiments and is shown on the schematic in figure~\ref{fig:schematic}.

Under the subduction mechanism, the size of the surface expression is determined by the position of the density front --- i.e. how far down the fjord the upper layer is displaced.
To estimate the displacement we consider that, in the absence of a subglacial discharge, the upper layer would occupy the full extent of the fjord.
The relaxation back to this steady state would be forced by the buoyancy difference between the upper layer and the surface expression fluid.
The velocity scale of such a flow is given by $u_f \sim \sqrt{g'_f \delta_f}$, where $g'_f$ is the reduced gravity based on the upper layer and surface expression densities and $\delta_f$ is the thickness of the surface expression or the upper layer, which have approximately the same values as in \citet{Mankoff16}.
We take $\delta_f=15\,$m based on figure 7 of \cite{Mankoff16}.
Values for $g'_f$ depend on the source volume flux and radius of the plume as they will impact the density of the surface expression.
However, as a representative case, using a radius of 10\,m and a source volume flux of $120\,{\rm m^3\,s^{-1}}$, gives $g'_f = 5.8\times10^{-3}\,{\rm m\,s^{-2}}$ and $u_f = 0.30\,{\rm m\,s^{-1}}$.
The velocity $u_f$ can be thought of as the velocity that the upper layer would propagate towards the ice face with, if the subglacial discharge were to be suddenly stopped.
However, the presence of the surface expression resists the flow of the upper layer towards the ice face and, at a steady state, the front between the upper layer and the surface expression will be stationary.
It follows that the position of this front will be the location where the surface expression velocity is equal to $u_f$.
If the surface expression velocity was greater than $u_f$ at the front then the upper layer would be pushed back downfjord. 
Alternatively, if the surface expression velocity was less than $u_f$ at the front then the upper layer would be able to propagate towards the ice face until a balance is obtained.

\begin{figure}
\centering
\noindent \includegraphics[width=19pc]{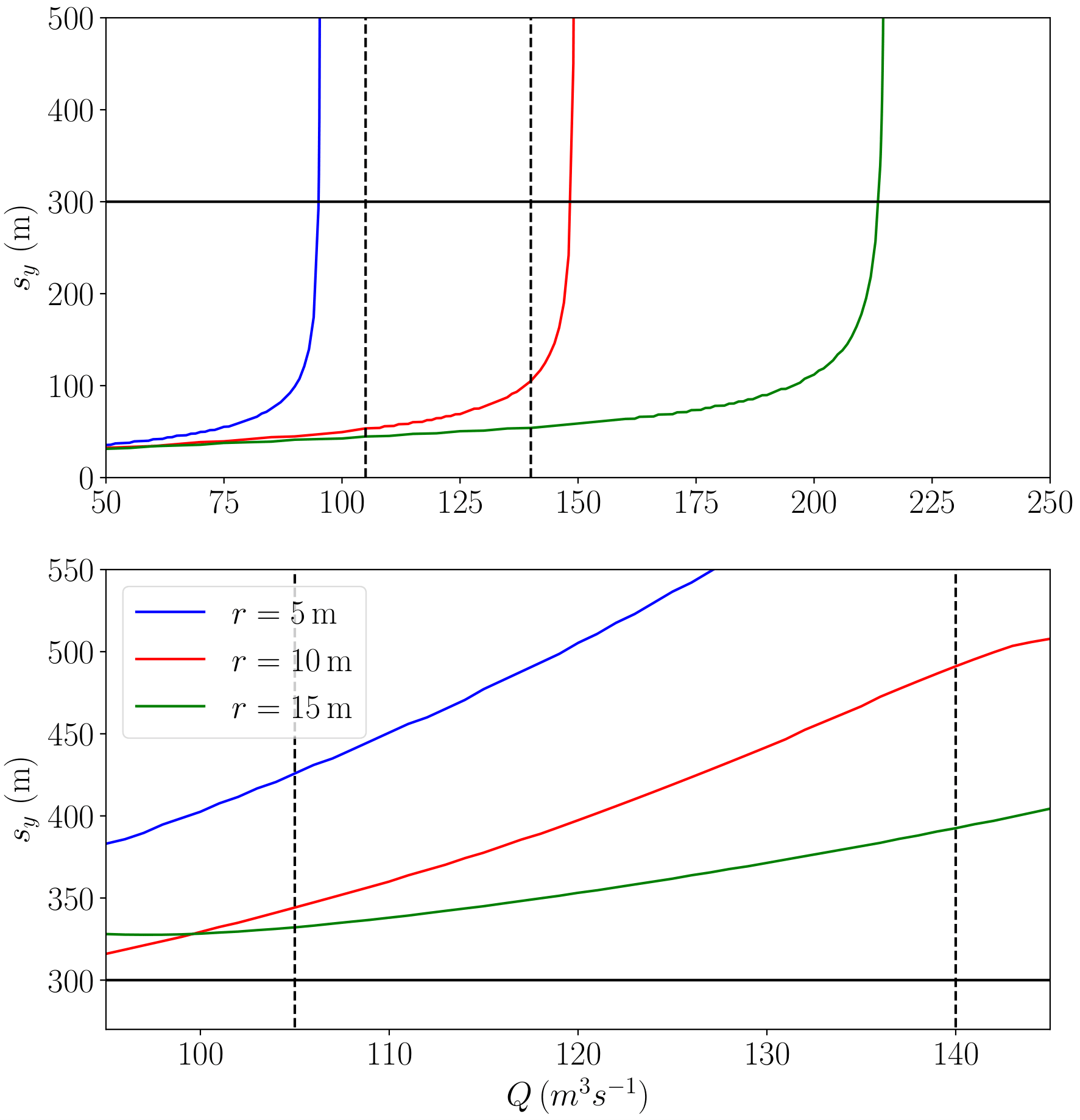}
  \caption{Calculation of the length of the subglacial discharge plume surface expression for a variety of source discharge volume fluxes and source radii.
  The top panel shows results using a two-layer stratification throughout the entire fjord and the \emph{sinking} mechanism.
  The bottom panel shows results using the stratification observed in \cite{Mankoff16} through the surface expression and the \emph{subduction} mechanism.
  Note the significantly different scales in both the $x$ and the $y$ axes for the two panels.
  Solid black lines show the observed length of the surface expression and dashed black lines show the estimated range of the subglacial discharge flux from \citet{Mankoff16}.}
  \label{fig:application}
\end{figure}

The model presented in \S\ref{sec:Theory} provides the velocity in the surface expression as a function of position.
As such we can use it to find where the surface expression velocity is equal to $u_f$ and use this location as an estimate of the surface expression size.
Figure~\ref{fig:application} shows estimates of the surface expression length as a function of discharge volume flux and discharge radius using both the sinking mechanism (top) and the subducting mechanism (bottom).
We note that although the sinking mechanism is not consistent with figures 5 and 7 of \cite{Mankoff16}, it is expected to be representative of the dynamics occurring when the subglacial discharge starts at the onset of the melting season.

Figure~\ref{fig:application} shows that the subduction mechanism predicts the size of the surface expression with more accuracy than the sinking mechanism.
When considering the sinking mechanism (top panel of figure~\ref{fig:application}), the surface expression is typically much shorter than that which was observed by \cite{Mankoff16}.
For the predicted surface expression size to be comparable to that which was observed, the density of the surface expression needs to be very similar to that of the upper layer, which leads to unrealistic sensitivity of the surface expression size to small changes in the discharge flow rate.
In contrast, the subduction mechanism predicts a surface expression size that is comparable to the observations, particularly for larger radii and lower source volume fluxes.
The lower end of the source volume flux range is also more consistent with the estimate based on RACMO data of $101\,{\rm m^3\,s^{-1}}$.

Combined with the observational evidence that vertical density profiles are stably stratified at all locations in the fjord \cite[figures 5 and 7 of][]{Mankoff16}, figure~\ref{fig:application} provides strong support for a subduction mechanism determining the surface expression length rather than a direct sinking mechanism.
We stress that subduction of the surface expression still relies on the presence of fluid that is less dense than the surface expression.
The difference is that for the surface expression to ``sink" the less dense fluid must be directly underneath the surface expression whereas if the less dense fluid is horizontally adjacent, the surface expression will be ``subducted".

It is worth considering why the difference between the laboratory experiments and observations exists.
The observational evidence suggests that the entire upper layer is displaced downfjord by the surface expression which is then subducted beneath the upper layer.
In contrast, the experiments clearly demonstrate sinking of the surface expression into a less dense underlying layer of fluid.
The key difference between the two situations is that in the fjord observations, the thickness of the surface expression is comparable to that of the upper layer, whereas in the laboratory experiments the upper layer is approximately ten times as thick as the surface expression.
When the surface expression and upper layer are of comparable thickness, the entire upper layer must be displaced downfjord and the interface below the surface expression remains stably stratified.
However, when the upper layer is significantly thicker than the surface expression, some portion of the upper layer remains below the surface expression and the interface becomes vertically unstable.

\section{Conclusion}
\label{sec:Conclusion}

We have considered the surface expression of a subglacial discharge plume in a stratified fjord. 
Several processes that could control the surface expression size and shape have been considered to attempt to understand the triangular surface expression observed at Saqqarliup Fjord \citep{Mankoff16}.
Our hypothesis is that if these processes can be better understood, subsurface properties of these plumes could be inferred from observations of the surface expression which are considerably simpler to make.

To this end, we have presented a model and experiments that examine the surface expression of a fountain released adjacent to a wall that is either vertical or sloping. 
The model separates the flow into vertical and horizontal regions. 
A transition region where all of the vertical momentum is converted to horizontal momentum through a free-surface pressure gradient connects the two regions. 
The model predicts that the dimensions of the surface expression of the fountain, non-dimensionalised by the thickness of the horizontal jet, will be linearly dependent on ${\rm{Fr}}$, which is consistent with previous studies of two-dimensional negatively buoyant surface jets \citep{Burridge17}.

Experiments are first used to examine the shape and spreading rate of the subsurface fountain. 
Similar to \cite{Ezhova18}, we find that the fountain spreads more rapidly in the wall-parallel direction than in the wall-perpendicular direction. 
Neither the rate of spreading nor the asymmetry are significantly affected by a sloping wall.

Experimental measurements of the dimensions of the surface expression generally agree with the model for a vertical wall. 
For a sloping wall, the model slightly under predicts the length of the surface expression.
At least for a vertical wall, the experiments appear to confirm that the transition from vertical to horizontal flow was treated appropriately in the model.
This provides some insight into how subglacial plumes transition to horizontal intrusions, which could help constrain the forcing in models of fjord-scale circulation.
However, a more complicated treatment of the transition is probably required if the ice face is significantly overcut.

The surface expression shape tends to become less semi-circular and more triangular as the size increases, consistent with the observed triangular surface expression at Saqqarliup Fjord \citep{Mankoff16}.
However, the predicted surface expression shape based on our model remains semi-elliptical rather becoming triangular (see, for example, figure~\ref{fig:plume_surface}).
This inconsistency highlights that, although the model can predict the shape and aspect ratio of the surface expression, it is not able to predict the shape of the surface expression at high values of ${\rm{Fr}}$ and requires further development.

Finally, the model is applied to observations of a subglacial discharge plume in front of Saqqarliup Sermia \citep{Mankoff16}. 
We apply the model in two separate ways exploring possible mechanisms by which the surface expression fluid could move away from the surface.
The first of these mechanisms involves \emph{sinking} into a less dense fluid body that underlies the surface expression.
The second mechanism involves the surface expression fluid being \emph{subducted} below a less dense fluid body that is horizontally adjacent.
Consistent with the oceanographic observations presented in \cite{Mankoff16}, predictions of the surface expression size are more accurate if the subduction mechanism is considered rather than the sinking mechanism.
The similarity between the model predictions and observations of the surface expression size suggests that it may be possible to infer the subglacial discharge properties at the source from observations of the downfjord density profile and observations of the surface expression size.
However, the experiments presented in this study were focused on the sinking mechanism and further experiments examining the subduction mechanism (i.e. with an upper layer of comparable depth to the surface expression) would help to demonstrate the applicability of the subduction mechanism.
In addition, comparison with observations from other fjords is needed to assess the generality of these results.
In particular, in a fjord where the surface expression is significantly less deep than the upper layer, the dynamics are expected to be different and, in this scenario, the sinking mechanism may be more appropriate.

%
\acknowledgments
We gratefully acknowledge technical assistance from Anders Jensen and thank anonymous reviewers for improving the clarity of the manuscript. 
CM thanks the Weston Howard Jr. Scholarship for funding. 
Support to CC was given by NSF project OCE-1434041 and OCE-1658079.

%






%
%
%

\appendix
\appendixtitle{Conversion of vertical momentum flux to horizontal momentum flux}\label{appA}
This appendix justifies the assumption made in \S\ref{sec:Theory}\ref{sec:transition} that the transition from vertical to horizontal flow can be treated as the flow around a $90^{\circ}$ corner with no loss of momentum or mass fluxes. 
The following calculation assumes incompressible and two-dimensional flow. 
Within the vertical plane through the centreline, and in the vicinity of the corner, these assumptions are expected to be valid.

We note that within this appendix $x$, $y$, $z$, and $w$ are used differently to the rest of the paper. 
Instead we use the standard nomenclature of complex variables used in \cite{Lamb16}.
The nomenclature within the appendix should be considered to be self-contained and independent from the remainder of the manuscript.

The inner boundary of the fountain is in contact with the stationary ambient fluid and, after subtracting the hydrostatic pressure, at high Reynolds number it can be regarded as a free surface (i.e. the pressure is zero). 
There is a general method using complex variables due to \cite{Lamb16}(\S\textbf{73)} which can be used for solving these problems in the inviscid, irrotational case. 
That is it conserves momentum and mass which should be a good approximation near the surface.

\begin{figure}
\centering
  \noindent \includegraphics[width=19pc]{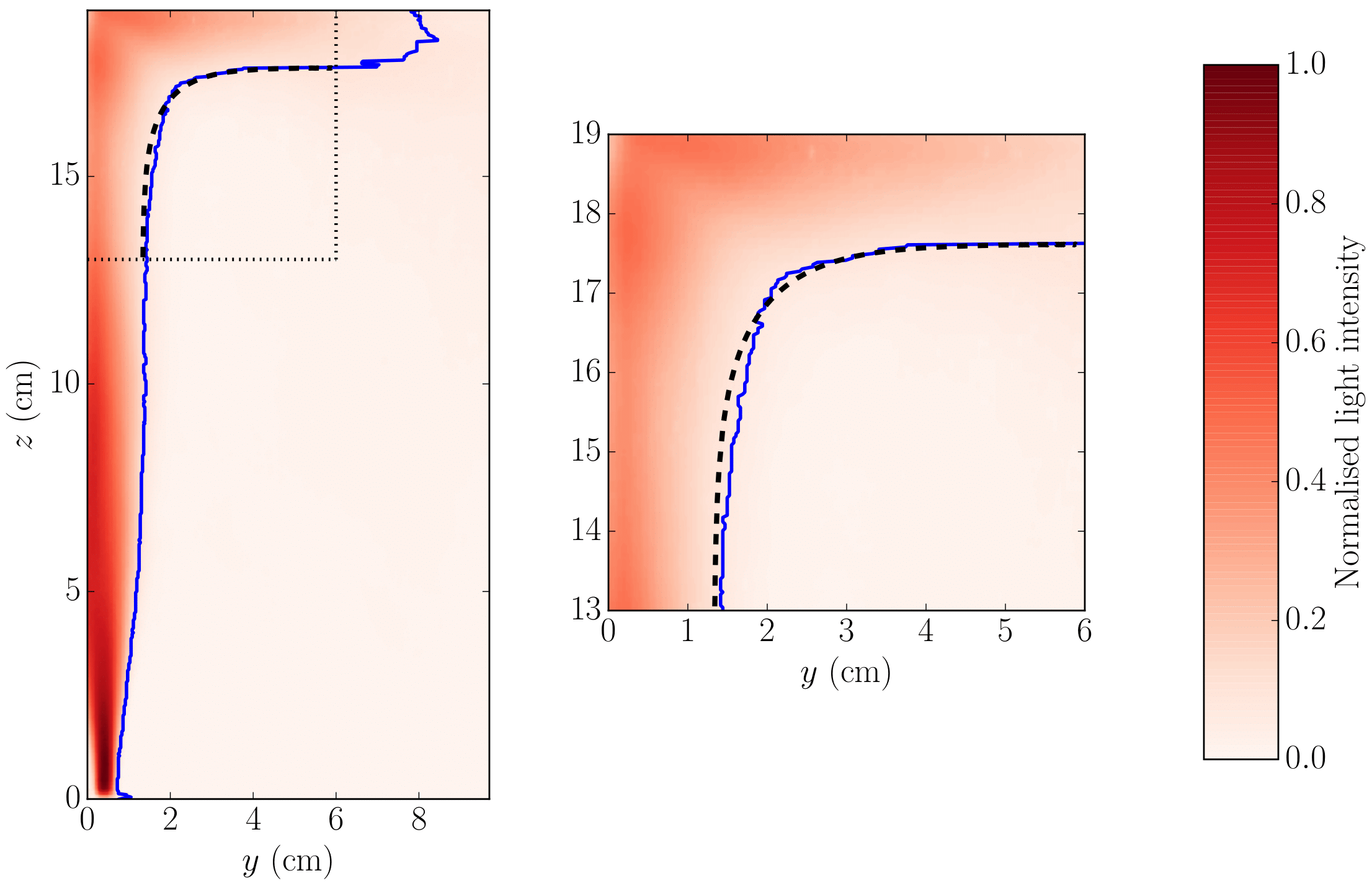}
  \appendcaption{A1}{Time-averaged normalised light intensity of a side view of the fountain 
  (i.e. $y$ and $z$ are the Cartesian coordinates). 
  The blue solid line shows where the light intensity falls to 20\% of the maximum intensity at that height which we use as a measure of the inner boundary of the fountain. 
  The black dashed line shows the solution for flow around a corner from (\ref{eq:xyi}) using Lamb's method. 
  Dotted lines show the enlarged section on the right panel.}
\label{fig:appendix}
\end{figure}

In our geometry there is a single right angle bend corresponding to a power of $1/2$, therefore following \cite{Lamb16} we must have
\begin{equation}
  \label{eq:dw}  
  \frac{{\rm{d}}z}{{\rm{d}}w}= \sqrt{\coth w},
\end{equation}
where $z=x+iy$ is the complex coordinate, 
$w = \phi + i\psi$, $\phi$ is the velocity potential 
and $\psi$ is the stream function. 
$\psi=0$ is the streamline along the wall and the horizontal free surface and $\psi=\pi/4$ is the streamline along the inner boundary of the fountain. 
Integrating (\ref{eq:dw}) gives
\begin{equation}
  \label{eq:1}
  z = \cot^{-1} \sqrt{\coth w}+\coth^{-1}\sqrt{\coth w},
\end{equation}
an implicit equation for the stream function $\psi$.
Applying Bernoulli's theorem the pressure is given by
\begin{equation}
  \label{eq:p}
  p = C -\frac{u^2+v^2}{2} = C -\frac{1}{2}\left|\frac{{\rm{d}}w}{{\rm{d}}z}\right|^2 = \frac{1}{2}-\frac{1}{2}|\tanh w|,
\end{equation}
where we have chosen the pressure to be zero at the inner boundary of the fountain.
The pressure along the vertical wall and the horizontal free surface is given by taking $w=s$, where $s$ is real and $w=0$ is the corner:
\begin{equation}
  \label{eq:pw}
  p_w = \frac{1}{1+e^{|2s|}}.
\end{equation}
The connection between $s$ and world coordinates comes from (\ref{eq:1}).

The equation for the inner boundary of the fountain is given implicitly by substituting $w=\pi/4+s$ into (\ref{eq:1}). 
In the vicinity of the bend we can write a parametric power series solution
\begin{eqnarray}
  \label{eq:xy}
  \begin{array}{lcl}x&=&\frac{\pi+\log(3+2 \sqrt2)}{4} \\[6pt]
  &&+ \frac{1}{\sqrt 2}\left(+s+\frac{1}{2} s^2-\frac{1}{6} s^3-\frac{5}{24} s^4 +\frac{17}{120} s^5\right),\end{array}\\
 \begin{array}{lcl}  y&=&\frac{\pi+\log(3+2 \sqrt2)}{4} \\[6pt]
  &&+ \frac{1}{\sqrt 2}\left(-s+\frac{1}{2} s^2+\frac{1}{6} s^3-\frac{5}{24} s^4 -\frac{17}{120} s^5\right).
  \end{array}
\end{eqnarray}
If we define $R=\left[\pi+\log(3+2 \sqrt2)\right]/2$ we can write this solution to the same order fully implicitly as
\begin{equation}
  \label{eq:xyi}
  R^2=(x+y)^2+(x-y)^2/\sqrt{2} +\frac{1}{8}\left(\frac{R}{\sqrt{18}}-1\right)(x-y)^4+\cdots.
\end{equation}
We can also look far from the corner where we have
\begin{equation}
  \label{eq:3}
  y = \frac{\pi}{4} + \exp(\pi/2-2x),
  \quad \mbox{or} \quad
  x = \frac{\pi}{4} + \exp(\pi/2-2y).
\end{equation}
Thus the solution corresponds to a jet of thickness $\pi/4$ and uniform incoming and outgoing speed $1$.  
The solution can trivially be scaled to any velocity and size.

Figure~A1 shows a time-averaged normalised light intensity image of the fountain viewed from the side of the tank with the solution for flow around a corner superimposed. 
The edge of the fountain is identified as the location where the normalised light intensity falls to less than 20\% of the maximum value at each height. 
The edge of the fountain is only used for comparison with the solution for flow around a corner and as such the exact threshold is not significant.

It can be seen that the solution given by (\ref{eq:xyi}) agrees well with the contour of light intensity near the free surface where the solution is expected to be valid, justifying the assumption of inviscid, irrotational flow in vicinity of the corner.

\bibliographystyle{ametsoc2014}
\bibliography{Surface_expression_ArXiV}

%

%

\end{document}